\documentclass{aa}
\usepackage{natbib}
\usepackage[varg]{txfonts}
\usepackage{graphicx}
\usepackage{float}

\usepackage{amsmath}

\usepackage{lscape}
\usepackage{rotating}
\usepackage{multirow}
\usepackage{hhline}
\usepackage{tabularx}
\usepackage{booktabs}
\usepackage{caption}

\newcolumntype{L}[1]{>{\raggedright\arraybackslash}p{#1}}
\newcolumntype{C}[1]{>{\centering\arraybackslash}p{#1}}
\newcolumntype{R}[1]{>{\raggedleft\arraybackslash}p{#1}}

\def\modif#1{\textcolor{black}{#1}}

\def\modtwo#1{\textcolor{black}{#1}}

\usepackage{color} 

\def\Msun{\ifmmode{\mathrm M_\odot}\else{M$_\odot$}\fi}

\newcommand{\HI}{\ion{H}{I}}

\begin{document}

\title{ALMA Resolves Giant Molecular Clouds in a Tidal Dwarf Galaxy}
\author{M.~Querejeta\inst{1}
\and F.~Lelli\inst{2}
\and E.~Schinnerer\inst{3}
\and D.~Colombo\inst{4}
\and U.~Lisenfeld\inst{5,6}
\and C.~G. Mundell\inst{7}
\and F.~Bigiel\inst{8}
\and S.~Garc\'ia-Burillo\inst{1}
\and C.~N.~Herrera\inst{9}
\and A.~Hughes\inst{10,11}
\and J.~M.~D.~Kruijssen\inst{12}
\and S.~E.~Meidt\inst{13}
\and T.~J.~T.~Moore\inst{14}
\and J.~Pety\inst{9,15}
\and A.~J.~Rigby\inst{2}
}

\institute{Observatorio Astron\'{o}mico Nacional, Alfonso XII, 3, E-28014 Madrid, Spain; \email{m.querejeta@oan.es}
\and
School of Physics and Astronomy, Cardiff University, 5 The Parade, CF24 3AA Cardiff, United Kingdom
\and
Max-Planck-Institut f\"ur Astronomie, K\"onigstuhl 17, D-69117 Heidelberg, Germany
\and
Max-Planck-Institut f\"ur Radioastronomie, Auf dem H\"ugel 9, D-53121 Bonn, Germany
\and
Departamento de F\'isica Te\'orica y del Cosmos, Universidad de Granada, E-18071 Granada, Spain
\and
Instituto Carlos I de F\'isica Te\'orica y
Computacional, Facultad de Ciencias, E-18071 Granada, Spain
\and
Department of Physics, University of Bath, Claverton Down, Bath, BA2 7AY, United Kingdom
\and
Argelander-Institut f\"ur Astronomie, Universit\"{a}t Bonn, Auf dem  H\"ugel 71, D-53121 Bonn, Germany
\and IRAM, 300 rue de la Piscine, F-38406 Saint Martin d’H\`eres, France
\and
Universit\'e de Toulouse, UPS-OMP, F-31028 Toulouse, France
\and
CNRS, IRAP, Av.\ du Colonel Roche BP 44346, F-31028 Toulouse cedex 4, France
\and
Astronomisches Rechen-Institut, Zentrum f\"{u}r Astronomie der Universit\"{a}t Heidelberg, M\"{o}nchhofstra\ss e 12-14, D-69120 Heidelberg, Germany
\and
Sterrenkundig Observatorium, Universiteit Gent, Krijgslaan 281 S9, B-9000 Gent, Belgium
\and
Astrophysics Research Institute, Liverpool John Moores University, IC2, Liverpool Science Park, 146 Brownlow Hill, Liverpool L3 5RF, United Kingdom
\and
Sorbonne Universit\'e, Observatoire de Paris, Universit\'e PSL, CNRS, LERMA, F-75005 Paris, France
}

\date{Received ..... / Accepted .....}

\abstract {
Tidal dwarf galaxies (TDGs) are gravitationally bound condensations of gas and stars formed during galaxy interactions. Here we present multi-configuration ALMA observations of J1023+1952, a TDG in the interacting system Arp~94, where we resolve \mbox{CO(2-1)} emission down to giant molecular clouds (GMCs) at $0.64'' \sim 45$\,pc resolution. We find a remarkably high fraction of extended molecular emission (${\sim}80{-}90$\%), which is filtered out by the interferometer and likely traces diffuse gas. We detect 111 GMCs that give a similar mass spectrum as those in the Milky Way and other nearby galaxies (a truncated power law with slope of $-1.76 \pm 0.13$). We also study Larson's laws over the available dynamic range of GMC properties ($\sim$2 dex in mass and $\sim$1 dex in size): GMCs follow the size-mass relation of the Milky Way, but their velocity dispersion is higher such that the size-linewidth and virial relations appear super-linear, deviating from the canonical values. The global molecular-to-atomic gas ratio is very high (${\sim}1$) while the \mbox{CO(2-1)/CO(1-0)} ratio is quite low (${\sim}0.5$), and both quantities vary from north to south.
Star formation is predominantly taking place in the south of the TDG, where we observe projected offsets between GMCs and young stellar clusters ranging from $\sim$50\,pc to $\sim$200\,pc; the largest offsets correspond to the oldest knots, as seen in other galaxies. In the quiescent north, we find more molecular clouds and a higher molecular-to-atomic gas ratio (${\sim}1.5$); atomic and diffuse molecular gas also have a higher velocity dispersion there. Overall, the organisation of the molecular ISM in this TDG is quite different from other types of galaxies on large scales, but the properties of GMCs seem fairly similar, pointing to near universality of the star-formation process on small scales.
}


\titlerunning{ALMA resolves GMCs in a Tidal Dwarf Galaxy}
\authorrunning{Querejeta et al.}

\maketitle 
\section{Introduction}
\label{Sec:introduction}

When galaxies collide, a portion of the gas is ejected by tidal forces and may eventually collapse under self-gravity, giving rise to new low-mass galaxies along the tidal debris.
These newborn systems, known as tidal dwarf galaxies (TDGs), are thought to be devoid of dark matter, but they retain the metallicity of the galaxy that they were originally stripped from \citep{2000AJ....120.1238D,2012ASSP...28..305D,2015A&A...584A.113L}. TDGs tend to have high gas fractions and they are often active star-formers \citep{2001A&A...378...51B,2016A&A...590A..92L}; many of them also host old stellar populations, but this is not always the case \citep{2016A&A...585A..79F}. Such a dark matter-free setting, subject to tidal forces from their neighbours, affords the possibility of testing theories of star formation in an extreme dynamical environment. While TDGs have been extensively mapped through deep optical imaging, only a few studies have looked at their molecular gas content and distribution, and these studies target spatial resolutions that are too coarse to resolve individual giant molecular clouds \citep[GMCs; e.g.][]{2000Natur.403..867B,2001A&A...378...51B,2007A&A...475..187D,2016A&A...590A..92L}.

One of the most fundamental challenges of modern astrophysics is understanding the process by which gas transforms into stars, and how this process is orchestrated as a function of environment.
Indeed, theories of galactic-scale star formation and cosmological simulations of galaxy formation are both guided by the Kennicutt-Schmidt relation \citep{1959ApJ...129..243S,1998ApJ...498..541K}, the observed correlation between the star formation rate surface density $\Sigma_{\mathrm{SFR}}$ and neutral gas surface density $\Sigma_{\mathrm{gas}}$ \citep[e.g.][]{2008MNRAS.383.1210S,2018ApJ...861....4S}. The correlation between star formation rates and the amount of molecular gas is also known to be very tight \citep[e.g.][]{2008AJ....136.2846B,2011AJ....142...37S,2013AJ....146...19L}.

An increasing body of observational work suggests that, when molecular gas is probed down to GMC scales, its properties depend on local physical conditions
\citep{2013ApJ...779...44H,2014ApJ...784....3C, 2017ApJ...839..133P}. Dynamical effects have been proposed to explain the differences in the state of molecular gas and its ability to form stars \citep[e.g.][]{2006ApJ...641..938L,2013ApJ...779...45M,2014MNRAS.440.3370K,2015MNRAS.454.3299R,2018ApJ...854..100M}. 
Cloud-cloud collisions have also been suggested as a trigger of star formation \citep[e.g.][]{2000ApJ...536..173T,2014MNRAS.445L..65F,2015MNRAS.446.3608D,2015ApJ...806....7T}.
Several other factors, like changes in the incident radiation field, might further contribute to the observed differences in the shapes of GMC mass spectra between M51, M33, and the Large Magellanic Cloud \citep{2013ApJ...779...46H}, or NGC\,6946, NGC\,628, and M101 \citep{2015ApJ...808...99R}. These environmental factors influence the ability of a GMC to form stars and, indirectly, the shape of the Kennicutt-Schmidt relation, but their relative importance remains unknown. To understand the physics of star formation, it is 
\modif{crucial} to probe conditions down to GMC scales, the units that are associated with massive star formation in our own Galaxy.

\subsection{Our Target: The Tidal Dwarf Galaxy J1023+1952}

\begin{figure*}[t]
\begin{center}
\includegraphics[trim=0 0 70 0, clip,width=0.95\textwidth]{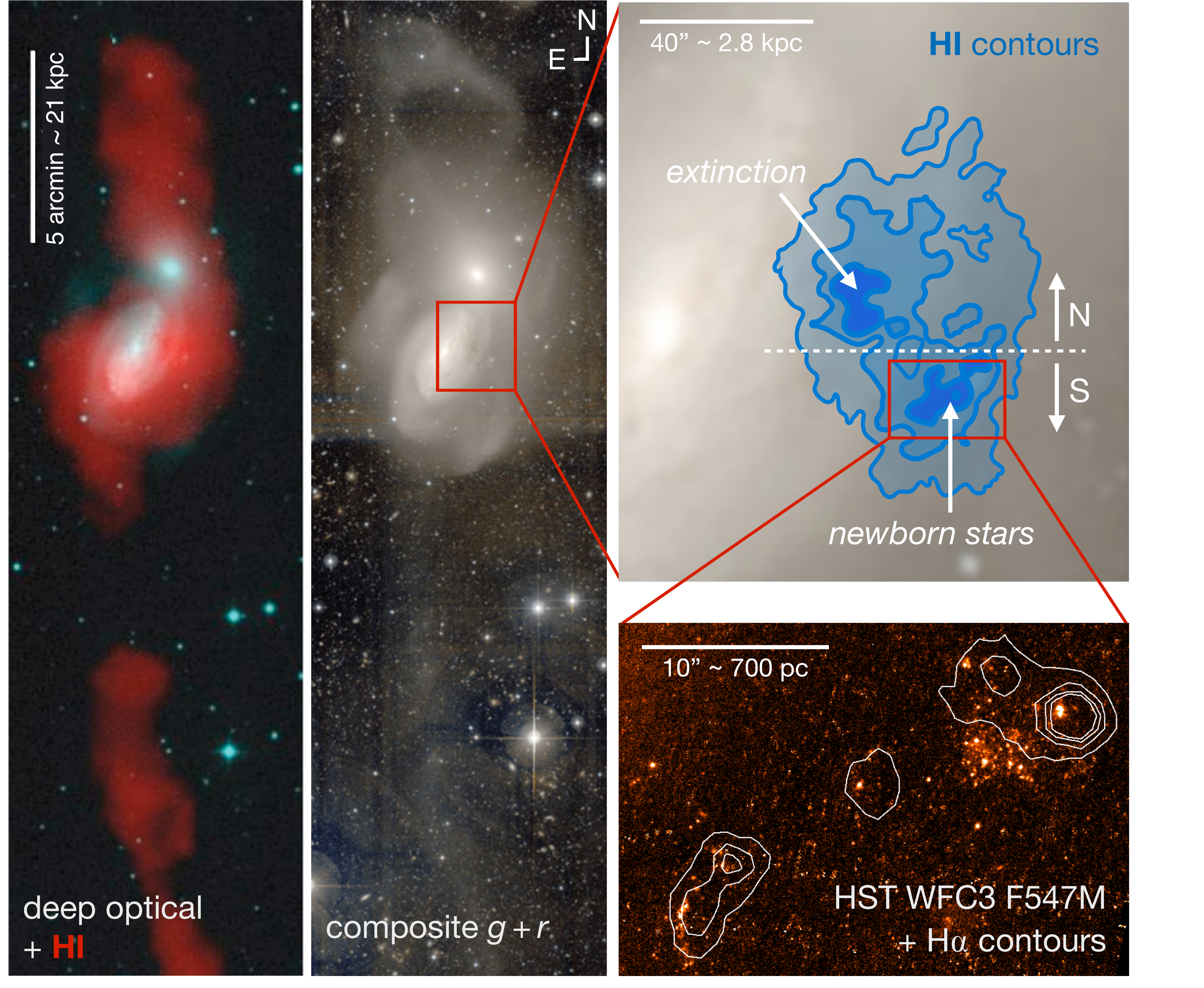}
\end{center}
\caption{\textit{Left panel:} \HI{} emission in the Arp\,94 system (in red) on top of a deep optical image from \citet{1995MNRAS.277..641M}. 
\textit{Central panel:} composite $g + r$ deep optical image from MegaCam/CFHT using an ``arcsinh stretch'' to emphasise faint emission \citep{2014MNRAS.440.1458D}. Both the left and central panels come from \citet{2014ApJ...797..117A}.
\textit{Top-right panel:} blow-up of the $g + r$ composite image showing the location of the TDG in \HI{} emission as blue contours, equivalent to an extinction of 0.4, 1.4, 2.4, and 3.4 A$\mathrm{_B}$ magnitudes\modif{, ranging from $4 \times 10^{20}$ to $4 \times 10^{21}$cm$^{-2}$} \citep{2004ApJ...614..648M}; the annotations indicate the area where a high optical extinction is seen towards the background galaxy (north-east) and the star-forming region in the south, as well as the demarcation line at  Dec${=}19^\circ 51' 50''$ that we apply to distinguish the northern from the southern part of the TDG. 
\textit{Bottom-right panel:} star-forming region in the south of the TDG as revealed by young clusters visible in HST WFC3/F547M imaging and H$\alpha$ emission from \citet{2004ApJ...614..648M} shown as contours (0.8, 2, 3, and $4 \times 10^{33}$~erg\,s$^{-1}$\,pc$^{-2}$).
}
\label{fig:mosaic}
\end{figure*}

J1023$+$1952 is a kinematically detached \HI{} cloud in front of the spiral galaxy NGC\,3227, in the interacting system Arp\,94 \citep{1995MNRAS.277..641M}.
Fig.\,1 shows our target in context, highlighting the large size of the Arp\,94 system and its complexity in terms of tidal tails as revealed by \HI{} and deep optical emission.
Specifically, the Arp\,94 system is composed of a spiral galaxy hosting a Seyfert 2 nucleus \citep[NGC\,3227;][]{1995MNRAS.275...67M,2001ApJ...549..254S,2014ApJ...794....2N} interacting with a ``green valley'' elliptical \citep[NGC\,3226;][]{2014ApJ...797..117A}.
The TDG lies at the base of the northern tidal tail and partially overlaps in projection with the disc of  NGC\,3227. \modif{Since the TDG} is obscuring stellar light from NGC\,3227, \modif{it} must be in front of the spiral galaxy \citep{1995MNRAS.277..641M,2004ApJ...614..648M}.
 The total content of atomic gas estimated by \citet{2004ApJ...614..648M} in the TDG is $M_\mathrm{HI} = 1.9 \times 10^8\,M_\odot$ (corrected for our assumed distance), distributed over a projected area of 7\,kpc~$\times$~5\,kpc. As shown in Fig.\,1, active star formation is ongoing in the southern half of the cloud, with blue knots clearly visible in H$\alpha$ emission at velocities that closely match those of the \HI{} cloud, ruling out the possibility that the star forming knots belong to the background galaxy, which is offset in velocity by almost 300\,km\,s$^{-1}$ \citep{2004ApJ...614..648M}. Furthermore, the metallicity of these star-forming knots is near-solar ($12+\log(\mathrm{O/H})=8.6$; \citealt{2008ApJ...685..181L}), comparable to the metallicity of the spiral galaxy NGC\,3227, and much higher than the metallicity that would be expected for a dwarf galaxy of this luminosity ($12+\log(\mathrm{O/H})\simeq8$, according to the magnitude-metallicity relation for classical dwarfs; \citealt{1995ApJ...445..642R,2010ApJ...710..663Z}). This strongly supports the idea that J1023+1952 is a TDG and not a pre-existing dwarf captured by the Arp\,94 system. 
 
 \modif{Using} the IRAM~30\,m telescope, \citet{2008ApJ...685..181L} demonstrated that this TDG is also rich in molecular gas ($M_\mathrm{H_2} = 1.5 \times 10^8\,M_\odot$, corrected for our distance). In spite of comparable atomic and molecular gas surface densities in the north and south of the cloud, star formation in J1023+1952 is predominantly happening along a 2-kpc-long ridge in the southern half (traced by optical blue colours, H$\alpha$, and UV emission, as well as young stellar clusters resolved with HST; see Fig.~1). 
 There might be some evidence for low-level star formation in the north-west of the TDG (see Sect.\,\ref{Sec:offsets}), but this is very limited compared to the vigorous star formation activity in the south. Thus, we distinguish two environments within the TDG: the quiescent north and the \modif{star-forming} south, separated by a declination cut at Dec${=}19^\circ 51' 50''$ (Fig.\,1). This cut is motivated by the two main concentrations of \HI{}, 
which are connected by a thin bridge on the eastern end.
It might also be possible that these two gas condensations correspond to two close-by but distinct TDGs, but the existing \HI{} data does not allow us to test this scenario by studying their internal dynamics.
The absence of detectable near-infrared continuum suggests a scarcity of old stars in the TDG \citep[$M^{\rm{old}}_{\star} < 4.7\times10^8$\,M$_\odot$ corrected for our distance;][]{2004ApJ...614..648M}, implying that this must be a young TDG which might not be in dynamical equilibrium. 

\subsection{Paper goals and structure}

TDGs constitute exciting laboratories to study how gas assembles within a newborn galaxy. In particular, it is yet not known how quickly gas condenses into compact structures in such a pristine system and whether those structures are similar to the GMCs observed in our Galaxy or in external galaxies.
To answer these questions, we present ALMA observations of J1023+1952 which resolve the molecular emission down to GMC scales. 

\modif{In Sect.\,\ref{Sec:data} we describe the ALMA data (\ref{Sec:ALMAdata}) and a rich set of ancillary data (\ref{Sec:ancillary}). In Sect.\,\ref{Sec:clouddecomp}, we identify and characterize the GMCs. In Sect.\,\ref{Sec:results} we present our results regarding the overall molecular structure (\ref{Sec:molecularstruc}), the molecular gas kinematics (\ref{Sec:kinematics}), and the GMC properties and scaling relations (\ref{Sec:GMC}). We discuss our results in Sect.\,\ref{Sec:Discussion} with a special focus on the connection between GMC properties, local environment, and star formation. We close the paper with a summary in Sect.\,\ref{Sec:conclusions}.}

We assume a distance of $14.5 \pm 0.6$\,Mpc \citep{2014ApJ...784L..11Y}, 
 based on a new AGN time lag method. As a consistency check, \citet{2014ApJ...784L..11Y} derived an independent value of $H_0 = (73 \pm 3)$\,km\,s$^{-1}$\,Mpc$^{-1}$ from these distances, in agreement with the literature \modif{\citep[e.g.][]{2020AJ....160...71S}}. For $d=14.5$\,Mpc, $1''$ corresponds to 70.3\,pc.
This is very similar to the $d=15.1$\,Mpc assumed by \citet{1995MNRAS.277..641M} based on the redshift of the source, and slightly smaller than
$d=20.4$\,Mpc assumed by \citet{2004ApJ...614..648M} and \citet{2008ApJ...685..181L} relying on the Tully-Fisher relation \citep{1988ngc..book.....T}.

\section{Data} 
\label{Sec:data}

\subsection{ALMA data: molecular gas emission} 
\label{Sec:ALMAdata}

We observed the \mbox{CO(2-1)} line with the ALMA Band 6 receiver (230\,GHz) during Cycle~4 (project 2016.1.00648.S, PI: M.~Querejeta). To recover emission from different spatial scales, we combined the 12\,m array with the Atacama Compact Array (ACA, i.e.\ 7\,m array and total power antennas). The 12\,m array observations were performed between 25 and 28 November 2016, with 46 antennas in C40-4 configuration, resulting in projected baselines between 15 and 704\,m which correspond to a typical angular resolution of $0.4{-}0.5''$ and a largest recoverable angular scale of ${\approx}4''$. The 7\,m observations were carried out between 4 and 26 October 2016 (with 8-10 antennas), and the total power, between 3 and 15 October 2016 (with 3-4 antennas). The 7m-array data at the \mbox{CO(2-1)} frequency provides a resolution of ${\approx}5.4''$ and a largest recoverable angular scale of ${\approx}29''$,
while ALMA total power has a resolution of ${\approx}28''$ and is by definition sensitive to emission from all spatial scales.

The full extent of the TDG (${\sim}64'' \times 43''$) was covered with an 18-pointing mosaic by the 12\,m array (primary beam of $\rm{FWHM} =  25.3''$); a 7-pointing mosaic was sufficient to cover the same field of view with the 7\,m array, given the larger primary beam ($\rm{FWHM} = 43.5''$). In both cases, the mosaic was hexagonally packed, with approximate Nyquist sampling along rows ($0.51 \times$ the primary beam) and a spacing $\sqrt{3}/2$ times the Nyquist separation between rows.

We calibrated and imaged the datasets using CASA 5.4.1 (Common Astronomy Software Applications\footnote{http://casa.nrao.edu}). After standard calibration, we concatenated the visibilities corresponding to the 7\,m and 12\,m interferometric data and imaged them together. We used the total power observations as model within \texttt{tclean} in CASA, so that the final cube recovers all the flux. The total power data were calibrated and imaged following the strategy presented in \citet{2020A&A...634A.121H}, Appendix~A. For the baseline correction, we fitted a polynomial of order 1 in a fixed velocity window (755 to 855\,km\,s$^{-1}$ and 1500 to 1600\,km\,s$^{-1}$). We matched the spatial grid and spectral resolution (2.5\,km\,s$^{-1}$ ) of the interferometric data.
We also performed an alternative imaging of the interferometric data, combining the visibilities from the 12\,m and 7\,m arrays (without total power).
In all cases, we cleaned using the Hogbom algorithm with natural weighting, down to a threshold of 2\,mJy per beam (${\sim}2\sigma$) in 2.5\,km\,s$^{-1}$ channels, and within the area where the primary beam response remains higher than 20\% of the maximum (which roughly corresponds to a tilted rectangle of ${\sim}80'' \times 70'' \approx 5.6 \times 4.9$\,kpc; shown as a dashed blue line in the left panel of Fig.\,\ref{fig:molec_frac}).
We chose a pixel size of $0.082''$ and an image size of $1344 \times 1440$ pixels (centred on RA${=}$10:23:26.249, Dec${=}+$19:51:58.83). With this imaging strategy, we obtained an average synthesised beam of $0.69" \times 0.60"$ ($\mathrm{PA} = 11^\circ$), which corresponds to ${\approx} 45$\,pc for our assumed distance.
The rms brightness sensitivity of the resulting data cube is $\sigma_{\rm rms} \sim 1$\,mJy per beam (${\approx} 56$\,mK), which for a representative linewidth of ${\rm FWHM} =15$\,km\,s$^{-1}$ yields a molecular gas surface density sensitivity of ${\sim} 3.9$\,$M_\odot\,{\rm pc}^{-2}$ (or a $1\sigma$ point-source sensitivity of ${\sim} 9 \times 10^3$\,$M_\odot$) for our assumed $\alpha_{\rm CO}$ conversion factor (Sect.\,\ref{Sec:transformtoSigmaGas}). In order to track flux recovery and compare against ancillary data, we also produced a version of the final cube convolved to lower resolution (with a circular Gaussian kernel of FWHM $= 1.5''$, $3''$, and $6.3''$, the latter to match the resolution of the \HI{} data).
We note that we did not detect any continuum emission. We applied a primary beam correction to the final cubes before performing any measurements. All velocities in this paper are heliocentric, expressed following the radio convention ($v= c (\nu_0 - \nu) /\nu_0$).

\subsection{Ancillary data} 
\label{Sec:ancillary}

\subsubsection{VLA \HI{} data} 
\label{Sec:VLA}

To trace atomic gas, we use \HI{} observations from the Very Large Array (VLA), originally published in \citet{2004ApJ...614..648M}. The data were taken by the VLA in B configuration, with a spatial resolution of $6.3''$ and velocity resolution of 10\,km~s$^{-1}$. Previously, \citet{1995MNRAS.277..641M} had used the VLA to map the same target at lower resolution (${\sim}20''$) in C configuration. The amount of diffuse \HI{} flux that is filtered out in B configuration is probably low, since the C array only recovered ${\sim}12$\% more flux than the B array \citep{2004ApJ...614..648M}.

\subsubsection{IRAM~30\,m data} 
\label{Sec:IRAM30m}

We compare our \mbox{CO(2-1)} observations against previous \mbox{CO(1-0)} observations of the TDG with the IRAM~30\,m telescope. Those single-dish observations were published by \citet{2008ApJ...685..181L}, and they attained a spatial resolution of $22''$.

\subsubsection{HST data} 
\label{Sec:HST}

We make use of archival data from the \textit{Hubble} Space Telescope (HST) taken with the Wide Field Camera 3 (WFC3). We obtained the data covering both NGC\,3227 and the TDG from the HST archive (proposal 11661, PI: Misty Bentz), which follows on-the-fly calibration procedures. The map was obtained with the medium-band filter F547M of the WFC3 camera and has been presented in \citet{2017ApJ...835..271B}. We corrected the astrometry of the HST image using Gaia field stars as reference \citep[a total of 15 stars down to 21\,mag in $g$-band, Gaia DR2;][]{2016A&A...595A...1G,2018A&A...616A...1G}. The offsets applied were $\Delta {\rm RA} = 1.2$\,px ($0.05''$) and $\Delta {\rm Dec} = - 7.8$\,px ($-0.31''$).

\subsubsection{H$\alpha$ data} 
\label{Sec:WHT}

We use H$\alpha$ observations of NGC\,3227 from the 4.2\,m William Herschel Telescope in La Palma, presented in \citet{2004ApJ...614..648M}. We corrected the astrometry of the H$\alpha$ image by applying an offset of $\Delta {\rm RA} = 0.33$\,px ($0.11''$) and $\Delta {\rm Dec} = 2.05$\,px ($0.67''$) to match Gaia field stars.

\subsection{Moment maps of the TDG emission}
\label{Sec:momentmaps}

As illustrated by Fig.\,1, the TDG partially overlaps in projection with the background galaxy NGC\,3227. To isolate the emission associated with the TDG from the emission arising from 
NGC\,3227, we followed a data-driven approach. Firstly, we smoothed the \HI{} cube to 30$''$ resolution in space and 50\,km~s$^{-1}$ in velocity, boosting the signal-to-noise ratio. Then, we clipped pixels below 3$\sigma_{\rm sm}$, where $\sigma_{\rm sm} = 1$\,mJy/beam is the rms noise per channel in the smoothed \HI{} cube. In this clipped cube, the \HI{} emission from the main galaxy and that from the TDG can be unambiguously identified because they cover distinct areas in space and velocity. Thus, we constructed a Boolean mask for the main galaxy, isolating contiguous areas in the channel maps. This mask was applied to the \HI{} and CO cubes at full resolution to exclude emission from NGC\,3227.

We obtained intensity maps for CO and \HI{} emission by integrating the corresponding cubes, \modif{after applying the mask excluding the emission from NGC\,3227}.
We integrated in the velocity range [1060, 1360] km\,s$^{-1}$. We consider the velocity range [1360, 1450] km\,s$^{-1}$ as line-free and use those channels to estimate the rms noise on a pixel-by-pixel basis. When computing moment maps, we applied a dilated mask technique to minimise the impact of noise \citep{2006PASP..118..590R}; we started from a threshold of $4\sigma$ based on the rms map, and then dilated the mask to include any adjacent voxels above $1.5\sigma$. 
We used the same dilated mask technique to construct first- and second-order moment maps. We obtained uncertainty maps for the first- and second-order moment maps by formally propagating $\mathrm{rms}_\mathrm{channel}$.

\subsection{Conversion to gas surface densities}
\label{Sec:transformtoSigmaGas}
  
To convert the ALMA \mbox{CO(2-1)} intensities into molecular gas surface densities, we apply a constant factor of $\alpha_\mathrm{CO}^{2-1} = 4.4\,M_\odot$\,(K\,km\,s$^{-1}$\,pc$^2)^{-1}$, which is the average value measured by \citet{2013ApJ...777....5S} directly on the \mbox{CO(2-1)} line for a set of resolved nearby galaxy discs (with a galaxy-to-galaxy scatter of 0.3\,dex, implying fluctuations of a factor of ${\sim}2$). The near-solar metallicity of the TDG suggests that this is probably a reasonable assumption. The constant $\alpha_\mathrm{CO}^{2-1}$ has the advantage that our surface density maps remain proportional to the directly measured \mbox{CO(2-1)} intensity. This value is equivalent to $2 \times 10^{20}$\,cm$^{-2}$\,(K\,km\,s$^{-1})^{-1}$ when applied on the \mbox{CO(2-1)} line, including a factor of 1.36 to correct for the presence of helium.

The standard Galactic value is 
$2 \times 10^{20}$\,cm$^{-2}$\,(K\,km\,s$^{-1})^{-1}$ when applied on the \mbox{CO(1-0)} line. 
\modif{Thus, considering the mean \mbox{CO(2-1)}/\mbox{CO(1-0)} line ratio of $0.5-0.6$ of this TDG (see Sect.\,\ref{Sec:R21}}),  
the fiducial molecular gas surface densities that we derive 
in this paper are ${\sim}40{-}50\%$ lower than what would be implied by the standard Galactic conversion factor applied on \mbox{CO(1-0)}.
In principle, we could also assume a spatially varying $\alpha_\mathrm{CO}^{2-1}$, assuming a constant $\alpha_\mathrm{CO}^{1-0}$ modulated by the observed \mbox{CO(2-1)}/\mbox{CO(1-0)} ratio map.
This would result in molecular gas surface densities that are a factor ${\sim}3$ and $\sim1.5$ higher in the north and south, \modif{respectively.}
We prefer to avoid this modulation of $\alpha_\mathrm{CO}$ by $R_{21}$ because of the coarse resolution ($\sim$28$''$) of the available $R_{21}$ map, limited by the \mbox{CO(1-0)} single-dish data.

We transformed \HI{} intensities into atomic gas surface densities using the standard formula for optically thin gas, $\eta_\mathrm{H}  = 1.82 \times 10^{18} \times F_\HI{}$, where $F_\HI{}$ is in K\,km\,s$^{-1}$ and $\eta_\mathrm{H}$ is in atoms per cm$^2$ \citep[e.g.][]{2016era..book.....C}; this is equivalent to multiplying by 0.0145 to transform from K\,km\,s$^{-1}$ to $M_\odot$\,pc$^{-2}$.

\section{GMC identification and characterisation}
\label{Sec:clouddecomp}

\subsection{GMC segmentation with CPROPS}
\label{Sec:CPROPS}

To perform the GMC identification with CPROPS \citep{2006PASP..118..590R}, we start from a mask of significance defined as those voxels with signal above $4\sigma$ expanded to any adjacent voxels with emission above $1.5\sigma$. The same mask of significance is used for SCIMES \citep{2015MNRAS.454.2067C}, as described in Appendix~\ref{Sec:SCIMES}.
Starting from the connected, discrete regions of signal within the mask (the so-called ``islands''), local maxima are subsequently identified and kept only if the voxels that are closer to a given maximum than to any other maxima define an area larger than the synthesised beam. If this area is more than one synthesised beam but less than two beams, the island is included in the catalogue but not decomposed. Additionally, the local maxima must be 2$\sigma_{\rm rms}$ above the merge level with other clouds. For all pairs of local maxima within an island, if the computed moments based on the emission associated with each of the maxima separately differs less than 100\% from the moments computed for the combined emission, then those are no longer considered as separate peaks. After this process, all the emission uniquely associated with a surviving local maximum is considered a segmentation unit (a GMC).

We consider emission as long as the primary beam response from ALMA is at least 20\% of the maximum, which corresponds to an area of ${\sim} 80'' \times 70''$.
We mask emission from the background galaxy NGC\,3227 as described in Sect.\,\ref{Sec:momentmaps}.
 
\begin{figure}[t]
\begin{center}
\includegraphics[trim=60 0 60 0, clip,width=0.45\textwidth]{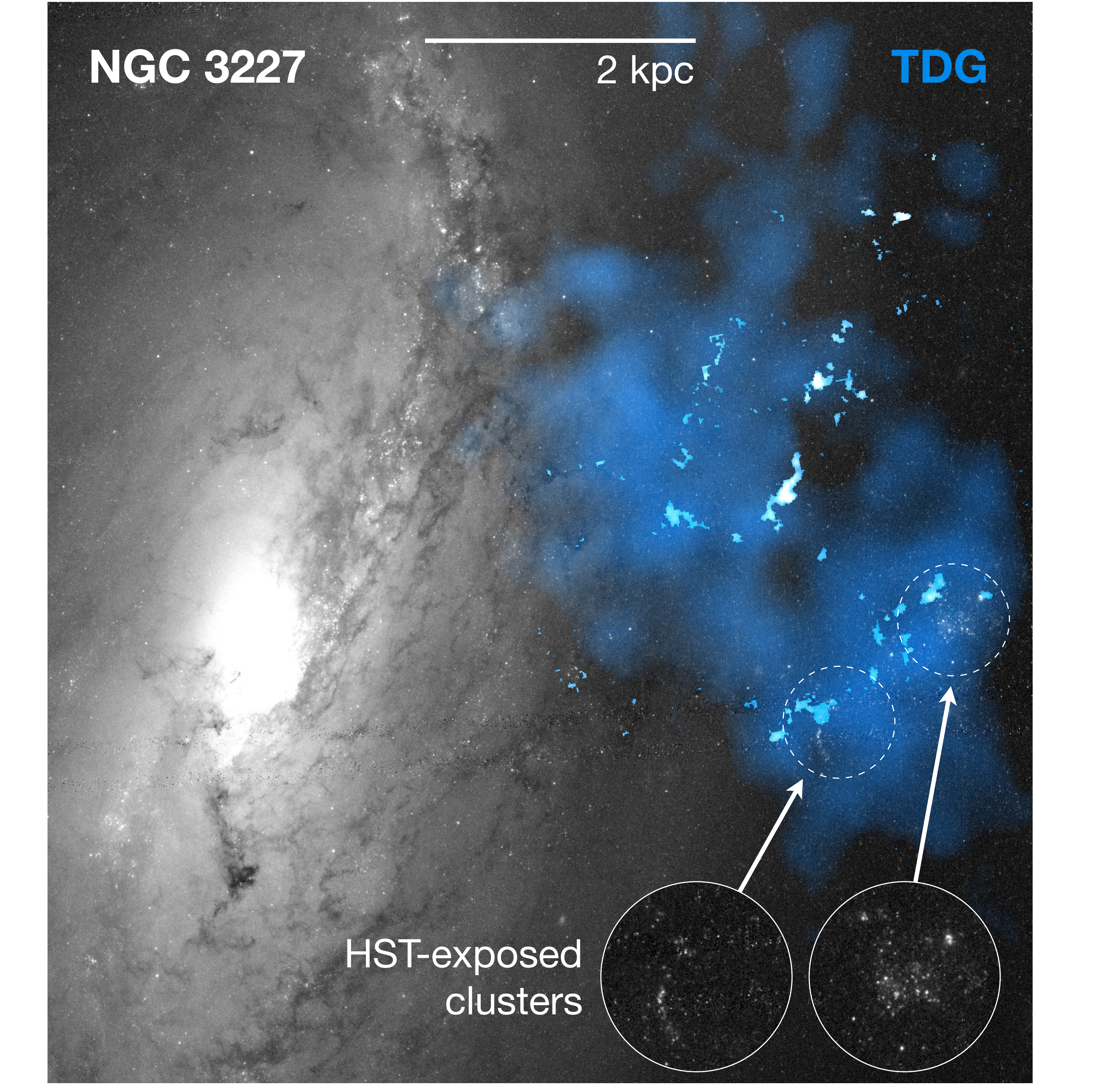}
\end{center}
\caption{False-colour image combining the \mbox{CO(2-1)} intensity \modif{from}
ALMA (white), optical image from HST (grayscale), and \HI{} intensity 
(blue) from the VLA \citep{2004ApJ...614..648M}. The circles 
show a blow-up where HST reveals young stellar clusters associated with the star-forming part of the TDG.
}
\label{fig:RGB}
\end{figure}

\begin{figure*}[t]
\begin{center}
\includegraphics[width=0.95\textwidth]{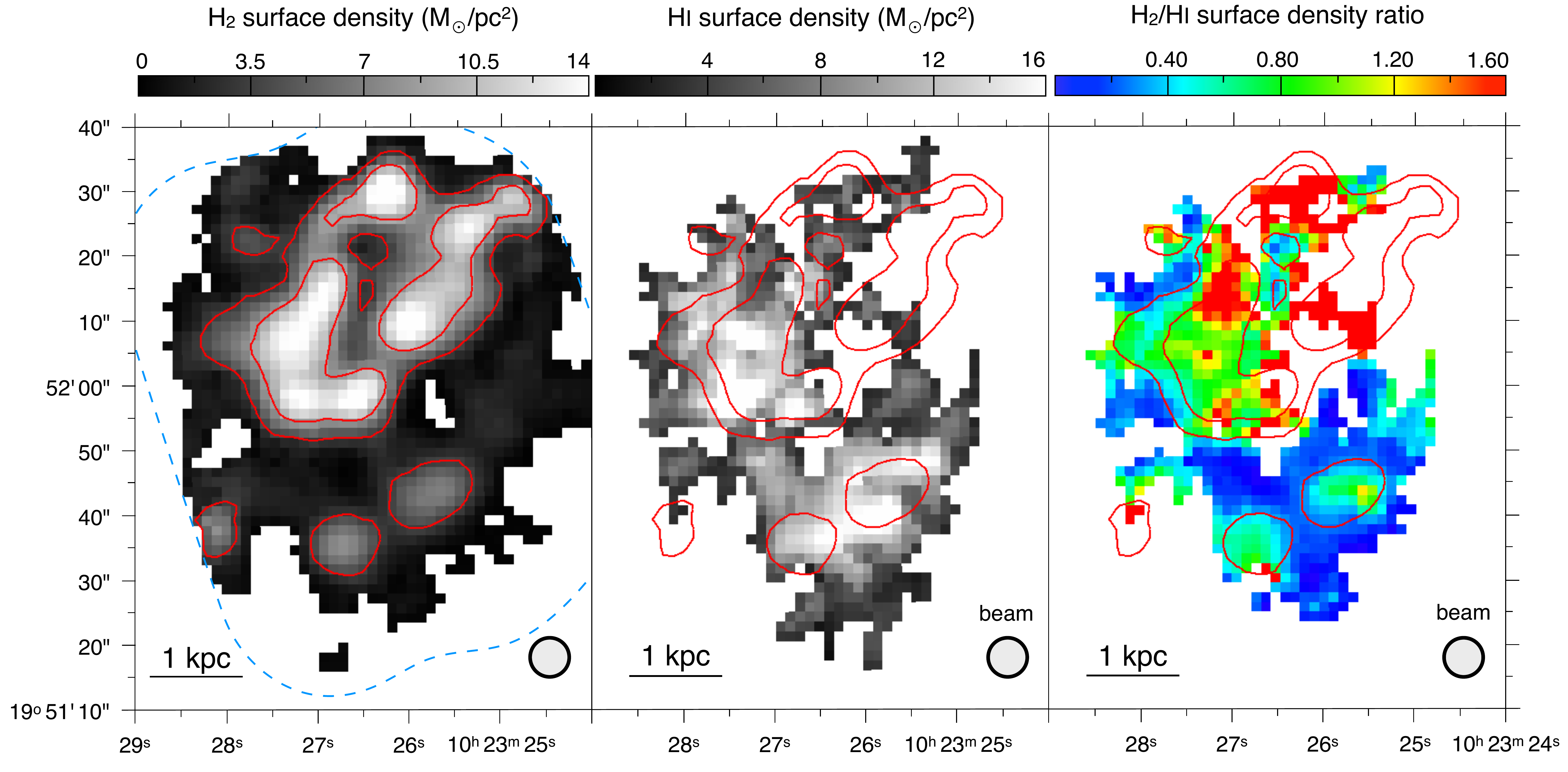}
\end{center}
\caption{\textit{Left panel:} H$_2$ surface density from ALMA \mbox{CO(2-1)} observations \modif{(12m\,+\,7m\,+\,TP)} convolved to $6.3''$ resolution, with contours highlighting levels of (2, 6, 10)\,M$_\odot$\,pc$^{-2}$. The dashed blue line is the ALMA field of view (where the primary beam response exceeds 20\%). \textit{Central panel:} \HI{} surface density from the VLA at $6.3''$ resolution \citep{2004ApJ...614..648M}, with CO contours from the left panel in red. \textit{Right panel:} the ratio of molecular-to-atomic gas surface densities derived from our \mbox{CO(2-1)} and \HI{} observations (left and central panel) at $6.3''$ resolution; CO contours as in the other two panels. In all maps north is up and east is left, with J2000.0 equatorial coordinates.
}
\label{fig:molec_frac}
\end{figure*}

\subsection{Deriving GMC properties}
\label{Sec:GMCproperties}

We follow \cite{2006PASP..118..590R} to extract physical properties out of the discrete units identified by CPROPS and SCIMES. The 
quantities extracted by the segmentation codes are the fluxes $F$, the sizes $R$, and the linewidths $\Delta V$.

Our ALMA cube can be characterised by a brightness temperature $T_{\rm i}$ for each volumetric pixel (voxel $i$). Each voxel can also be identified by two spatial coordinates ($x_{\rm i}$, $y_{\rm i}$) and a velocity channel ($v_{\rm i}$); according to the segmentation, a given GMC covers a discrete range of pixels and velocity channels ($\delta x$, $\delta y$, $\delta v$). The luminosity is obtained adding up all emission within the desired range of voxels:

\begin{equation}
L_{\rm CO}=\sum_{\rm i} T_{\rm i}\,\delta x\,\delta y\,\delta v\,D^2,
\end{equation}

\noindent
where $D$ is the distance to the galaxy. To obtain sizes, \modif{both CPROPS and SCIMES first rotate} the $x$ and $y$ axes to be aligned with the major and minor axes of the cloud, determined from principal component analysis. Subsequently, the size of the cloud is calculated as the intensity-weighted second moment ($\sigma_{\rm maj}$, $\sigma_{\rm min}$) along each spatial dimension:

\begin{equation}
\sigma_{\rm maj} = \sqrt{\frac{\sum_{\rm i} T_{\rm i}\,(x_{\rm i} - \bar{x})^2}{\sum_{\rm i} T_{\rm i}}}, ~ \sigma_{\rm min} = \sqrt{\frac{\sum_{\rm i} T_{\rm i}\,(y_{\rm i} - \bar{y})^2}{\sum_{\rm i} T_{\rm i}}},
\end{equation}

\noindent
with $\sigma_{\rm r} = \sqrt{\sigma_{\rm maj} \sigma_{\rm min}}$. The size of the cloud, $R$, is calculated as $R=\eta \sigma_{\rm r} $, where $\eta$ translates the spatial second moment to the radius of a spherical cloud, and it depends on the density distribution of the GMC. We adopt $\eta = 1.91$ \citep{1987ApJ...319..730S}, which is the most standard assumption in Galactic and extragalactic studies.
When the cloud size provided by the segmentation codes is smaller than the synthesized beam, \modif{the GMC is considered unresolved and is excluded from the analysis.}

Analogously, the velocity dispersion is obtained as:

\begin{equation}
\sigma_{\rm v} = \sqrt{\frac{\sum_{\rm i} T_{\rm i}\,(v_{\rm i} - \bar{v})^2}{\sum_{\rm i} T_{\rm i}}},
\end{equation}

\noindent
with the FWHM linewidth given by $\Delta v = \sqrt{8 \ln(2)} \sigma_{\rm v}$.

Starting from the basic properties above, we also derive a number of indirect properties, including luminous and virial masses. The luminous mass is obtained as $M_{\rm lum} = \alpha_\mathrm{CO}^{2-1} L_{\rm CO}$, while the virial mass is estimated as $M_{\rm vir} = 1040 \sigma_v^2 R$ ($M_{\rm vir}$ in M$_\odot$ if $\sigma_v$ is in km\,s$^{-1}$ and $R$ in pc), which assumes clouds with density profile $\rho \propto R^{-1}$.

To account for the effects of finite sensitivity, which can yield slightly smaller GMC sizes than in reality, \modif{we used the extrapolation of the moments to the 0\,K contour implemented in CPROPS and SCIMES \citep{2006PASP..118..590R}.} Additionally, the synthesised beam was deconvolved from the extrapolated spatial moments. To account for uncertainties, we ran CPROPS with the bootstrap option, performing 1000 iterations.

\section{Results}
\label{Sec:results}

Figure~\ref{fig:RGB} shows the distribution of the molecular gas revealed by ALMA (white) on top of atomic gas (blue) and stellar continuum (grayscale). This composite-colour image already hints at some of the most important results of this paper. The \HI{} observations at $6.3''$ resolution show that atomic gas extends over a relatively large projected area (${\sim}6$\,kpc$ \times 4$\,kpc$\,= 24$\,kpc$^2$). The ALMA observations at high-resolution ($0.64''$) reveal that the molecular emission that stands out on sub-arcsecond scales is highly clumpy and has a low covering factor. However, a very large fraction of the molecular gas is diffuse; we examine this in Sect.\,\ref{Sec:molecularstruc}, 
\modif{together with}
the molecular-to-atomic gas ratio and \mbox{CO(2-1)/CO(1-0)} ratio.
We study the GMC properties, mass spectra, and GMC scaling relations in Sect.\,\ref{Sec:GMC}. In Sect.\,\ref{Sec:kinematics} we focus on gas kinematics. The clusters visible in Fig.\,\ref{fig:RGB} (highlighted by the circular blow-ups) are often not aligned with the peaks of CO emission; we look at these offsets in Sect.\,\ref{Sec:offsets}.

\subsection{Molecular structure of the TDG}
\label{Sec:molecularstruc}

\subsubsection{Molecular-to-atomic gas ratio}
\label{Sec:molecfrac}

The left and central panels of Fig.\,\ref{fig:molec_frac} show the CO and \HI{} intensity maps at matched $6.3''$ resolution, overlaid with CO contours. 
There is plenty of CO emission where \HI{} emission is very limited (below our detection threshold of $4\sigma$), particularly towards the north-west of the TDG. The map also suggests that CO emission is somewhat clumpier than \HI{} emission.
Indeed, the standard deviation in the flux distribution in the CO map (0.45 on a logarithmic scale) is almost twice higher than \HI{} (0.25), confirming that \HI{} is more homogeneous even at matched resolution.
These effects contribute to the spatial variation of the molecular-to-atomic gas ratio evident in the right panel of Fig.\,\ref{fig:molec_frac}: there are important changes in the H$_2$/\HI{} ratio both locally (around CO peaks) and globally (between the north and south of the TDG).

\begin{table*}[t!]
\begin{center}
\caption[h!]{Molecular and atomic \modif{gas properties}
at matched $6.3''$ resolution.}
\begin{tabular}{lccc}
\hline\hline 
		&	 Whole TDG &	 North$^a$ & South$^b$	\\
\hline 
Total H$_2$ mass (M$_\odot$)  &	$8.6 \times 10^7$ & $7.3 \times 10^7$	&	$1.3 \times 10^7$ \\
Total \HI{} mass (M$_\odot$) & $8.4 \times 10^7$ & $4.9 \times 10^7$	&	$3.5 \times 10^7$ \\
 \hline
 Mean $\Sigma_{\rm H2}$ (M$_\odot$\,pc$^{-2}$)  &	4.8	& 5.7	& 2.5 \\
Mean $\Sigma_{\HI{}}$  (M$_\odot$\,pc$^{-2}$) & 8.1 	& 7.8 	& 8.4 \\
 \hline
  Total H$_2$ mass / Total \HI{} mass  &	1.02	& 1.47	& 0.37 \\
  Mean $\Sigma_{\rm H2} / \Sigma_{\HI{}}$ & 0.59 &  0.73 &  0.30 \\
 \hline
   Mean $R_{21} = $CO(2-1)/CO(1-0) & 0.52 & 0.56 & 0.38 \\
 \hline
\end{tabular}
\label{table:molfrac}
\end{center}
\tablefoot{(a) North is where Dec${>}19^\circ 51' 50''$. (b) South is where Dec${<}19^\circ 51' 50''$. \modif{Measurements at matched spatial resolution ($6.3''$) applying a mask of significance to both cubes ($4\sigma$ dilated mask); thus, the total H$_2$ mass quoted here is lower than implied by ALMA at native resolution without any masking ($1.6 \times 10^8$\,M$_\odot$).} The mean $\Sigma_{\rm H2}$ and $\Sigma_{\HI{}}$ are calculated for pixels with significant detections of $H_2$ and \HI{}, respectively; this means that they are averaged over different areas.}
\end{table*}

Table~\ref{table:molfrac} lists our measurements of the molecular-to-atomic gas ratio for the two environments in the TDG. \modif{Similar to other TDGs \citep{2001A&A...378...51B},} this system is highly molecular with a total H$_2$/\HI{} mass ratio of ${\sim}1$. However, 
the molecular ratio varies significantly across the TDG \modif{with local variations up to a factor of 2-3}. 
In addition to local variations, globally the quiescent north has a substantially higher molecular-to-atomic gas ratio (1.47) than the star-forming south (0.37). 
\modif{This becomes extreme in the north-west corner, where the $\Sigma_{\rm H2} / \Sigma_{\HI{}}$ ratio is at least} ${\sim}4{-}6$ considering $4\sigma$ upper limits for \HI{}.
\modif{Averaging} the local $\Sigma_{\rm H2}$ to $\Sigma_{\HI{}}$ ratio for pixels with simultaneous CO and \HI{} detections results in a lower ratio than dividing the \modif{total H$_2$ and \HI{} masses}.
 Independently from the indicator used, the molecular-to-atomic gas ratio is \modif{surprisingly} higher in the quiescent north, whereas star formation is enhanced in the more atomic gas-dominated south. The average H$_2$ surface density is also higher in the north, while the average \HI{} surface density is similar among both environments. The molecular-to-atomic ratio remains higher in the north \modif{even if we use} a spatially variable $\alpha_\mathrm{CO}^{2-1}$ (Sect.\,\ref{Sec:transformtoSigmaGas}), but the difference between north and south becomes smaller.

\subsubsection{Diffuse molecular gas}
\label{Sec:diffusefrac}

\begin{figure*}[t]
\begin{center}
\includegraphics[width=0.95\textwidth]{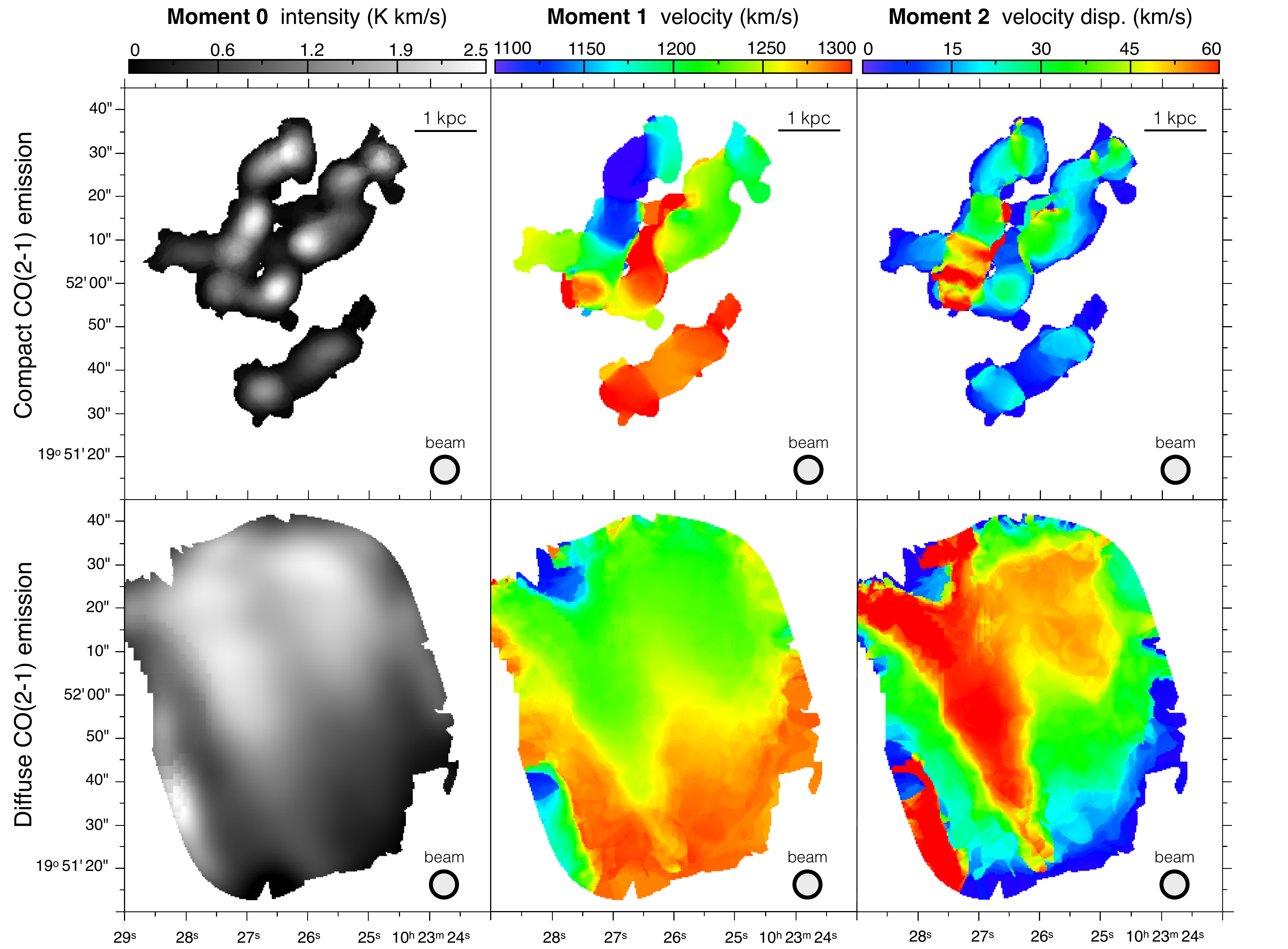}
\end{center}
\caption{\textit{Top panels:} Compact emission traced by the ALMA 12m\,$+$\,7m arrays without TP.  \textit{Bottom panels:} Diffuse molecular emission obtained by subtracting the compact emission channel by channel (top panels) from the total molecular emission measured by ALMA 12m\,$+$\,7m\,$+$\,TP (i.e.\ including short spacings): this is the emission that is filtered out by the interferometer. \textit{Left panels:} zeroth-order moment maps,
measuring the integrated intensity (in K km\,s$^{-1}$). \textit{Middle panels:} first-order moment maps,
representing the velocity field (in km\,s$^{-1}$). \textit{Right panels:} second-order moment maps,
\modif{tracing CO linewidths ($\sigma$ in} 
km\,s$^{-1}$).
In all panels, the grey circle in the bottom right represents the FWHM of the synthesised beam ($6.3''$).}
\label{fig:diffuse_gas}
\end{figure*}

\begin{table*}[t!]
\begin{center}
\caption[h!]{Fluxes (in K\,km\,s$^{-1}$\,kpc$^2$) retrieved with or without short spacings and in GMCs at the native ALMA resolution ($0.64''$).}
\begin{tabular}{lccccc}
\hline\hline 
\multirow{2}{*}{Environment} & \multirow{2}{*}{\begin{tabular}{c} Area \\ (kpc$^2$) \end{tabular}} & \multicolumn{2}{c}{Flux in 12m+7m+TP} & \multirow{2}{*}{\begin{tabular}{c} Flux in 12m+7m  \\ inside mask$^d$ \end{tabular}} & \multirow{2}{*}{Flux in GMCs$^e$}	\\
     \cmidrule(lr){3-4}
		&	  & total  & inside mask$^d$ &	  & 	\\
\hline 
Whole TDG$^a$ & 25.6 & 37.1 & 4.86 (13.1\%) & 4.58 (94.2\%) & 6.76 (18.2\%) \\
North$^b$ &     15.8 & 28.5	& 3.93 (13.8\%) & 3.70 (94.1\%) & 5.44 (19.1\%) \\
South$^c$ &      9.9 & 8.5	& 0.93 (10.9\%) & 0.88 (94.6\%)  & 1.32 (15.5\%) \\
 \hline
\end{tabular}
\label{table:fluxes}
\end{center}
\tablefoot{(a) Whole TDG is the entire ALMA field-of-view, where the primary beam response exceeds 20\% (dashed blue line in Fig.\,\ref{fig:molec_frac}; ${\sim} 80'' \times 70''$). (b) North is where Dec${>}19^\circ 51' 50''$. (c) South is where Dec${<}19^\circ 51' 50''$. (d) Flux inside 4$\sigma$ dilated mask (the mask is identical in both cases and is based on the 12m+7m cube). The fourth and fifth columns indicate the percentage of the total 12m+7m+TP flux that is inside the dilated mask and the percentage of 12m+7m+TP flux recovered with 12m+7m inside the same mask, respectively. (e) Total flux in all GMCs identified with CPROPS, after extrapolation to 0\,K and deconvolution of the beam; without extrapolation and deconvolution, the total flux in GMCs is considerably lower, 2.46\,K\,km\,s$^{-1}$\,kpc$^2$. The percentage of the total 12m+7m+TP flux is listed. The total flux in GMCs identified with SCIMES (Appendix\,\ref{Sec:SCIMES}) is 8.18 and 2.94\,K\,km\,s$^{-1}$\,kpc$^2$ for the extrapolated and non-extrapolated version, respectively.}
\end{table*}

 \modif{Table\,\,\ref{table:fluxes} provides the CO flux recovered by our observations using different ALMA arrays. Specifically,} the 12m array (configuration C40-4) and the compact 7m array are sensitive to scales up to $4'' \sim 280$\,pc and $29''\sim 2$\,kpc, respectively, while the combination of 12m\,+\,7m\,+\,total power should capture emission from all scales.
Given the very low covering factor of emission in the 12m\,+\,7m cube, we apply a mask to ensure that we extract only meaningful flux (using a 4$\sigma$ dilated mask as explained in Sect.\,\ref{Sec:momentmaps}). This mask encapsulates the compact structures that the 12m\,+\,7m interferometer is sensitive to. We measure the flux inside the same mask both for the 12m\,+\,7m and 12m\,+\,7m\,+\,TP cubes:
the 12m\,+\,7m cube recovers almost the whole flux (94\%) in these compact structures. However, if we compare this compact emission against the total CO flux in the 12m\,+\,7m\,+\,TP cube, almost 90\% of the total CO flux is outside the mask.

Extended emission can also be quantified as the fraction of the total CO flux that is contained in the GMCs identified by CPROPS (see Sect.\,\ref{Sec:CPROPS}). Depending on whether we include the extrapolation to infinite sensitivity or not, the flux in GMCs is 18\% or 7\% of the total, respectively. This roughly agrees with the amount of flux in compact structures that is captured by the interferometer (12m\,+\,7m data). 
\modif{Thus, most} molecular emission is not arising from the type of compact structures that are responsible for massive star formation in galaxies. The extended emission is likely associated with diffuse molecular gas, and it accounts for as much as ${\sim}80{-}90$\% of the molecular emission in the TDG. This is substantially higher than the values found in other nearby galaxies \citep[typically 10-60\%;][]{2013ApJ...779...43P,2015AJ....149...76C}, only comparable to the $74{-}91$\% extended emission found in the bar of NGC\,1300 \citep{2020MNRAS.495.3840M}. There is no strong dependence of the diffuse fraction on environment: it is slightly higher in the north but quite similar in both cases (see Table\,\ref{table:fluxes}).

Figure~\ref{fig:diffuse_gas} shows the spatial distribution of compact CO emission at $6.3''$ resolution (ALMA 12\,m$+$7\,m data, without total power) against the diffuse component (the result of subtracting the intensity from ALMA 12\,m$+$7\,m$+$TP minus the intensity from 12\,m$+$7\,m only, channel by channel). \modif{We perform this analysis at $6.3''$ resolution to avoid being dominated by local spikes due to noise, and because it is the resolution of the \HI{} map.} As expected, the diffuse component is smoother than the compact emission even at matched $6.3''$ resolution. Yet, the diffuse emission is not perfectly homogeneous either, and seems to be more intense towards the north.

The diffuse component shows a north-south velocity gradient and, as one might expect, the velocity in diffuse CO varies more smoothly than in compact CO structures.
The CO linewidth is high ($\sigma \sim 50{-}60$\,km\,s$^{-1}$), and correlates with intensity.
This is in stark contrast with the compact component, which shows large local variations in velocity and linewidth, as well as much more limited velocity dispersion (mostly below ${\sim} 20{-}30$\,km\,s$^{-1}$). The kinematics of the compact component does not follow in detail the velocity field and dispersion map of the diffuse component.

\begin{figure}[t]
\begin{center}
\includegraphics[trim=50 0 40 0, clip,width=0.48\textwidth]{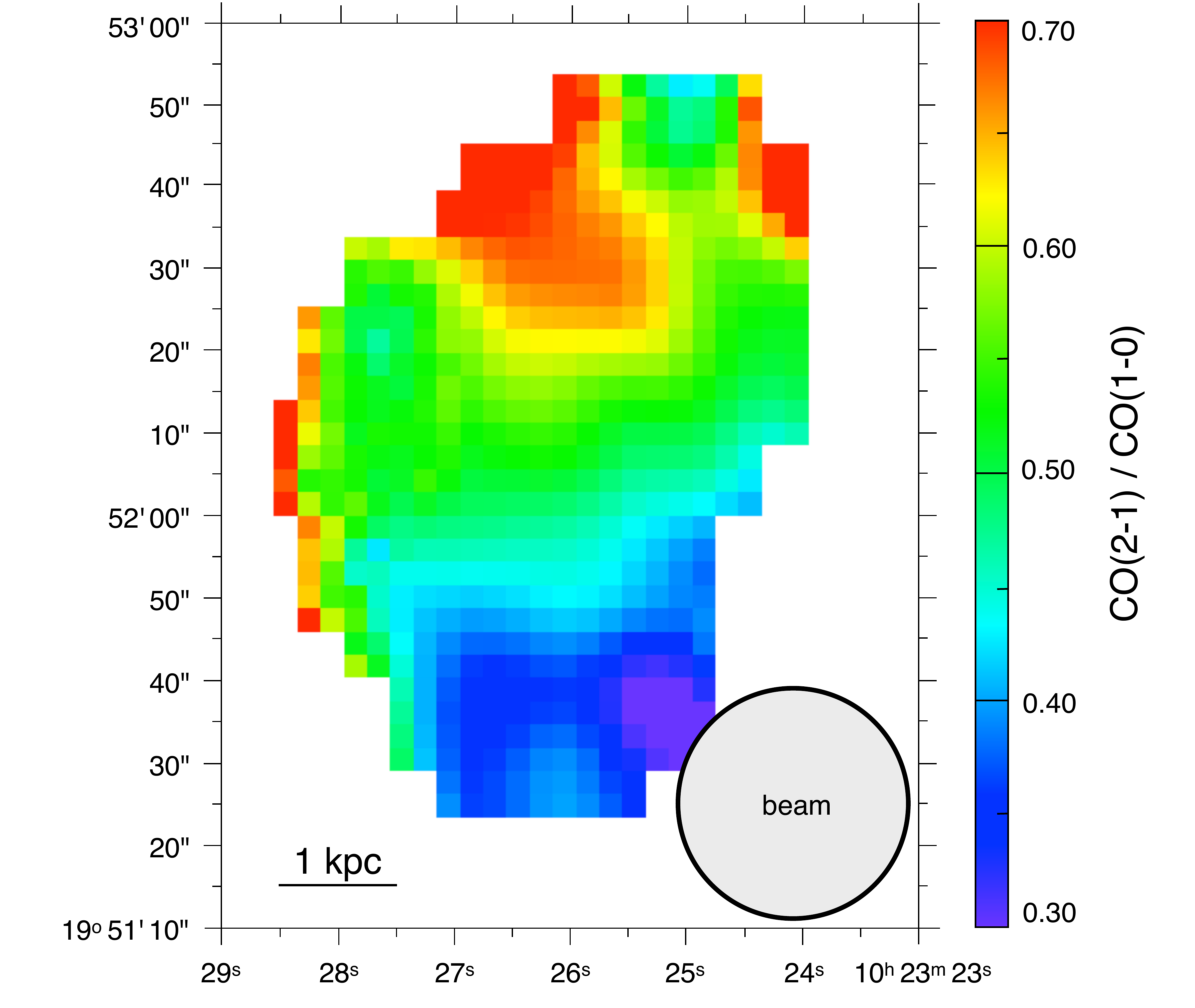}
\end{center}
\caption{CO(2-1)/CO(1-0) ratio at $28''$ resolution.
}
\label{fig:R21}
\end{figure}

\subsubsection{CO(2-1)/CO(1-0) ratio}
\label{Sec:R21}

We explore spatial variations in $R_{21}$ by combining our ALMA total power \mbox{CO(2-1)} observations ($28''$ resolution) with the IRAM~30\,m \mbox{CO(1-0)} single-dish observations ($22''$ resolution) from \citet{2008ApJ...685..181L}. We first match resolutions by convolving the IRAM~30\,m cube to $28''$ resolution. Then, we compute the relevant moment-zero maps (Sect.\,\ref{Sec:momentmaps}) and take their ratio,  \mbox{CO(2-1)/CO(1-0)}, in flux units of K\,km\,s$^{-1}$. The average value (masking NGC\,3227) is 0.52, in good agreement with the global $R_{21} = 0.54 \pm 0.10$ found by \citet{2008ApJ...685..181L}.

Fig.\,\ref{fig:R21} shows that there is a significant gradient in $R_{21}$ along the TDG, from a maximum of ${\sim} 0.7$ in the north down to a minimum of ${\sim} 0.3$ in the south.
This might indicate changes in the physical properties of molecular gas between the quiescent and star-forming part of the TDG\modif{. We also} compute $R_{21}$ for the part of the background galaxy NGC\,3227 simultaneously covered by the ALMA and IRAM 30\,m observations. In NGC\,3227, we find an average value of 0.53, similar to the global ratio for the TDG. In Sect.\,\ref{Sec:discussR21} we discuss how the mean $R_{21}$ ratio in the TDG is compatible with the lowest values observed in nearby \modif{galaxies.}

\begin{figure*}[t]
\begin{center}
\includegraphics[width=0.95\textwidth]{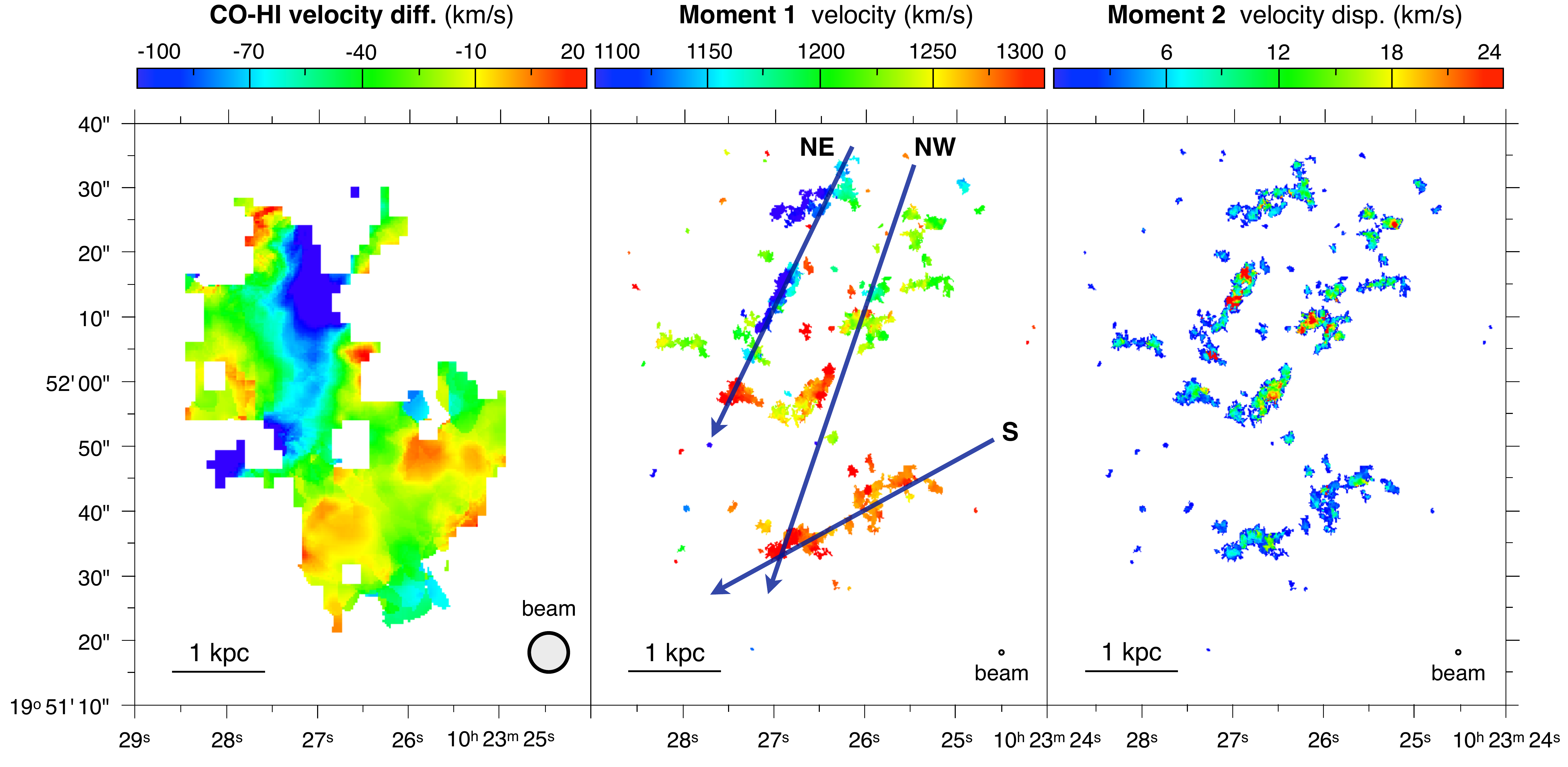}
\end{center}
\caption{\textit{Left panel:} difference between moment-1 maps for CO (ALMA 12\,m$+$7\,m$+$TP) and \HI{} at matched $6.3''$ resolution. \textit{Middle panel:} CO velocity field at $0.64'' \approx 45$\,pc resolution measured as the moment-1 map from ALMA (12\,m$+$7\,m$+$TP), using a 4$\sigma$ dilated mask. The blue arrows are the slits used for the position-velocity diagrams from Fig.\,\ref{fig:pvdiagram}. \textit{Right panel:} CO velocity dispersion measured as the moment-2 map from ALMA (12\,m$+$7\,m$+$TP), using the same 4$\sigma$ dilated mask.}
\label{fig:mom1mom2}
\end{figure*}

\begin{figure*}[t]
\begin{center}
\includegraphics[trim=0 34 0 10, clip,width=1.0\textwidth]{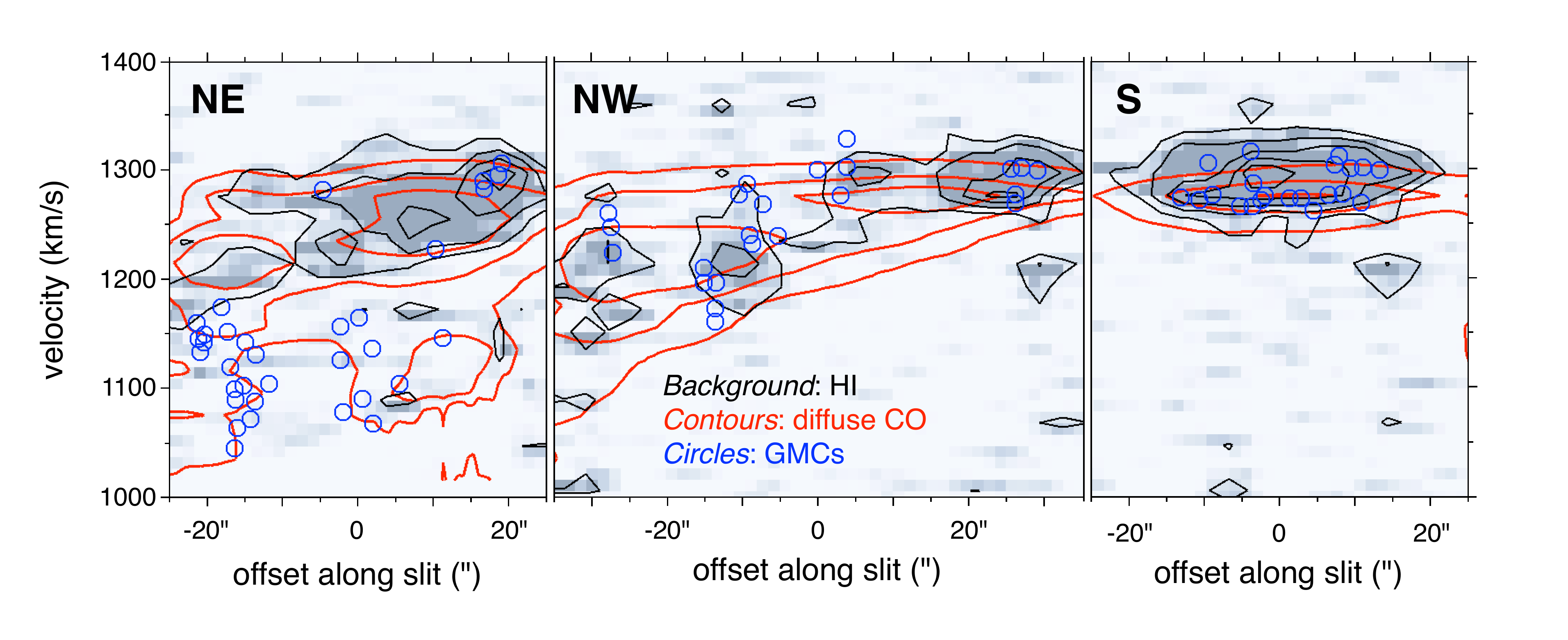}
\end{center}
\caption{
Position-velocity diagrams of \HI{} (at $6.3''$ resolution) in grayscale and black contours along the three slits indicated in Fig.\,\ref{fig:mom1mom2}. Red contours show the distribution of diffuse CO emission from ALMA (12\,m$+$7\,m$+$TP minus 12\,m$+$7\,m only), \modif{corresponding to 6, 9, 12\,mK}. Blue circles indicate the location of GMCs identified with CPROPS.}
\label{fig:pvdiagram}
\end{figure*}

\subsection{Kinematics}
\label{Sec:kinematics}

\modif{The \HI{} emission} shows a clear velocity gradient along the north-south direction which was interpreted by \citet{2004ApJ...614..648M} as possibly due to rotation.
The left panel of Fig.\,\ref{fig:mom1mom2} shows the difference between the moment-1 maps for \HI{} and CO at matched $6.3''$ resolution. There are significant velocity offsets towards some regions in the north (up to ${\sim}100$\,km\,s$^{-1}$).
Conversely, there is reasonably good agreement between CO and \HI{} velocities towards the south (mostly within $\pm 20$\,km\,s$^{-1}$).

\modif{The middle and left panel of Fig.\,\ref{fig:mom1mom2} focus on the highest resolution available from ALMA. These high-resolution moment maps show differences with \HI{} since} 
the CO peaks seem to be organised into several kinematically coherent filaments. 
Rather than a smooth gradient, there are quite abrupt changes in the velocities of these \modif{filaments}. This could be indicating that the \modif{high-density ISM}
is made up of relatively independent sub-structures, \modif{embedded} in a more diffuse molecular medium.

Figure\,\ref{fig:pvdiagram} shows position-velocity diagrams of \HI{} along the three slits indicated in Fig.\,\ref{fig:mom1mom2}. Simultaneously, we track as red contours the diffuse CO emission 
\modif{at a matched resolution of $6.3''$}
(i.e.\ the emission that is filtered out by the interferometer, like the bottom panels of Fig.\,\ref{fig:diffuse_gas}). 
To first-order approximation, there is very good agreement between the atomic gas 
and diffuse CO, and they both reflect the north-south velocity gradient seen at lower resolution. There are two noteworthy differences, \modif{though:} 1) in the NE ridge, there is plenty of diffuse CO at low velocities (${\sim}1050{-}1150$\,km\,s$^{-1}$) where the VLA cube did not register significant \HI{} emission; 2) there is an offset of ${\sim}30$\,km\,s$^{-1}$ between \HI{} and diffuse CO in the star-forming ridge in the south. 

We can also examine the distribution of GMCs (Sect.\,\ref{Sec:clouddecomp}) on the position-velocity diagrams of Fig.\,\ref{fig:pvdiagram}. 
\modif{GMCs} tend to cluster around regions of enhanced \HI{} emission\modif{, except in} the NE ridge, where most GMCs are offset from \HI{} by ${\sim}50{-}150$\,km\,s$^{-1}$ towards lower velocities. There is diffuse CO emission coinciding with or near those offset GMCs, but almost no significant \HI{} visible \modif{(some diffuse, faint \HI{} emission is however visible in a deeper cube from the C-array of the VLA; \citealt{1995MNRAS.277..641M})}. 
This offset towards lower velocities is suggestive of an overlap of gaseous structures along the line of sight: the distribution of diffuse emission and GMCs is rather continuous, starting from the main \HI{} emission around ${\sim}1200{-}1300$\,km\,s$^{-1}$ and extending towards velocities closer to the disc of NGC\,3227 (${\sim}1000$\,km\,s$^{-1}$). 
This may indicate that a gaseous bridge connects the TDG with the background spiral galaxy. The clustering of clouds at lower velocities can be well visualised through a rotating position-position-velocity representation of the GMCs which is accessible as supplementary material\footnote{https://bit.ly/2M6f6Tg}.

Another important issue from the kinematic point of view is the higher CO velocity dispersion found in the north of the TDG compared to the south. \citet{2008ApJ...685..181L} suggested that this might reflect an increase in large-scale turbulence in the north that might, in turn, prevent gas from forming stars in that part of the TDG. 
Taking the velocity centroid from each GMC (Sect.\,\ref{Sec:GMC}), we compute the statistical dispersion of those velocities for the ensemble of clouds in the north and in the south.  Within GMCs, the intrinsic velocity dispersions are slightly higher in the north (median $\sigma_{\rm GMC} \sim 6.5$\,km\,s$^{-1}$) than in the south ($\sigma_{\rm GMC} \sim 4.5$\,km\,s$^{-1}$), and the relative cloud-to-cloud velocity dispersion is also higher in the north ($\sigma_{\rm cloud-cloud} \sim 73$\,km\,s$^{-1}$) than in the south ($\sigma_{\rm cloud-cloud} \sim 46$\,km\,s$^{-1}$ and as low as 16\,km\,s$^{-1}$ if restricted to the star-forming area). 

\subsection{GMC properties and scaling relations}
\label{Sec:GMC}

The exquisite resolution of ALMA ($0.64'' \approx 45$\,pc) allowed us to resolve large and intermediate-size GMCs in this system, with molecular gas masses ranging from ${\sim} 10^4$\,M$_\odot$ to ${\sim} 10^6$\,M$_\odot$. We estimate our $5\sigma$ completeness limit as $5 \times 10^4$\,M$_\odot$ based on the point-source sensitivity derived in Sect.\,\ref{Sec:ALMAdata}.

We employed the CPROPS algorithm to identify GMCs \citep{2006PASP..118..590R}, as described in Sect.\,\ref{Sec:CPROPS}. In the Appendix\,\ref{Sec:SCIMES} we also present an alternative approach using SCIMES \citep{2015MNRAS.454.2067C}, a dendrogram-based method \citep{2008ApJ...679.1338R} which, on top of segmenting the molecular ISM, informs on how it is hierarchically structured. In both cases, we follow the same strategy to derive physical properties (Sect.\,\ref{Sec:GMCproperties}). The results for the cloud ensemble derived using CPROPS and SCIMES are very similar, and that is why we focus on CPROPS next, while we point the interested reader to the Appendix\,\ref{Sec:SCIMES} for the details on SCIMES. The final catalogue of GMC properties from CPROPS is presented in Table\,\ref{table:results}.

\begin{figure}[t]
\begin{center}
\includegraphics[trim=50 0 90 0, clip,width=0.45\textwidth]{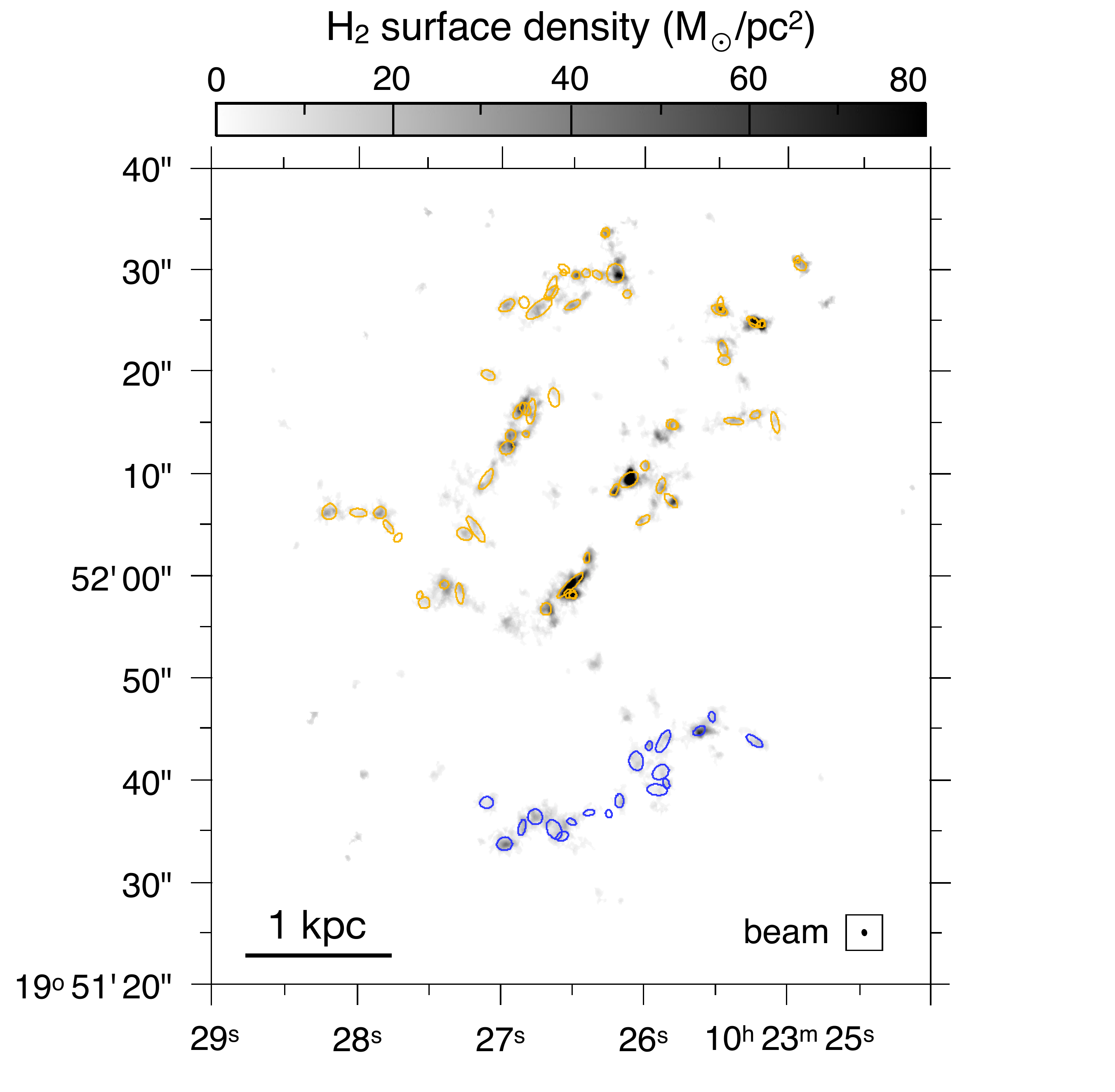}
\end{center}
\caption{GMC positions and orientations indicated on the ALMA \mbox{CO(2-1)} integrated intensity map. GMCs are shown as ellipses with the CPROPS extrapolated and deconvolved major and minor axes (2nd moment of emission). Orange and blue ellipses represent clouds in the north and south of the TDG, respectively. The bottom-right corner shows the ALMA synthesised beam ($0.69'' \times 0.60''$ with PA=11$^\circ$).
}
\label{fig:GMC_map}
\end{figure}

\begin{table*}[t!]
\begin{center}
\caption[h!]{GMC Properties as derived by CPROPS.}
\setlength{\tabcolsep}{5pt} 
\begin{tabular}{lcccccccccc}
\hline\hline
\multirow{2}{*}{Property} & \multirow{2}{*}{Unit}  & \multicolumn{3}{c}{TDG}  & \multicolumn{3}{c}{north} & \multicolumn{3}{c}{south} \\
     \cmidrule(lr){3-5} \cmidrule(lr){6-8} \cmidrule(lr){9-11}
& & min  & median & max  & min  & median & max & min  & median & max \\
   \hline
$T_{\rm max}$ & K &  0.3 &  0.4 &  1.0 &  0.3 &  0.4 &  0.8 &  0.3 &  0.5 &  1.0 \\
$L_{\rm CO}$ & K\,km\,s$^{-1}$\,pc$^{2}$ &   1.8$\times 10^3$ &   3.9$\times 10^4$ &   4.5$\times 10^5$ &   1.8$\times 10^3$ &   4.1$\times 10^4$ &   4.5$\times 10^5$ &   3.7$\times 10^3$ &   3.5$\times 10^4$ &   1.6$\times 10^5$ \\
$R$ & pc &   12.3 &   58.0 &  123.4 &   12.3 &   59.7 &  123.4 &   15.8 &   58.0 &  102.5 \\
$\sigma_v$ & km\,s$^{-1}$ &    0.9 &    5.8 &   19.7 &    0.9 &    6.6 &   19.7 &    0.9 &    4.5 &    9.1 \\
$M_{\rm lum}$ & $\mathrm{M_\odot}$ &   8.1$\times 10^3$ &   1.7$\times 10^5$ &   2.0$\times 10^6$ &   8.1$\times 10^3$ &   1.8$\times 10^5$ &   2.0$\times 10^6$ &   1.6$\times 10^4$ &   1.5$\times 10^5$ &   7.0$\times 10^5$ \\
$M_{\rm vir}$ & $\mathrm{M_\odot}$ &   2.3$\times 10^4$ &   2.7$\times 10^6$ &   4.0$\times 10^7$ &   2.8$\times 10^4$ &   3.4$\times 10^6$ &   4.0$\times 10^7$ &   2.3$\times 10^4$ &   2.3$\times 10^6$ &   7.8$\times 10^6$ \\
  \hline
\end{tabular}
\label{table:GMCproperties}
\end{center}
\end{table*}

\subsubsection{GMC properties}
\label{Sec:GMCstats}

We find a total of 111 GMCs with CPROPS (out of which 81 have extrapolated and deconvolved values), and Fig.\,\ref{fig:GMC_map} shows their distribution, sizes, and orientation. Far from clustering mostly towards the star-forming south (19 clouds), the number of GMCs is larger in the north (62 clouds), even though the projected area in the north is only ${\sim}50$\% larger. As expected, the distribution of GMCs follows the areas of strongest CO emission.
Some GMCs are rounder (axis ratio ${\sim} 1$) while others are more elongated (${\sim} 3$), and reassuringly we find no correlation between the orientation of the GMCs and the PA of the ALMA synthesised beam.

Table~\ref{table:GMCproperties} summarises the typical properties of the GMCs. We find GMCs with radii of a few tens of parsec (up to ${\sim}100$\,pc), luminous masses ranging ${\sim}10^4$ to ${\sim}10^6$\,M$_\odot$, and velocity dispersions of a few km\,s$^{-1}$. These values are comparable to massive GMCs in the Milky Way and other nearby galaxies \citep[e.g.][]{2008ApJ...686..948B,2009ApJ...699.1092H,2014ApJ...784....3C,2018ApJ...857...19F}, in spite of the starkly different nature of the TDG. Virial masses are 
larger than luminous masses, as we will discuss later.

We find no significant differences between global properties of GMCs in the north and south of the TDG other than ${\sim}50$\% higher $\sigma_v$ and $M_\mathrm{vir}$ \modif{in the north} (Table~\ref{table:GMCproperties}). We confirmed that these results do not change qualitatively if we rely on the segmentation from SCIMES instead of CPROPS (Table~\ref{table:GMCproperties}).

\subsubsection{GMC mass spectra}
\label{Sec:GMCMS}

\begin{figure}[t]
\begin{center}
\includegraphics[width=0.5\textwidth]{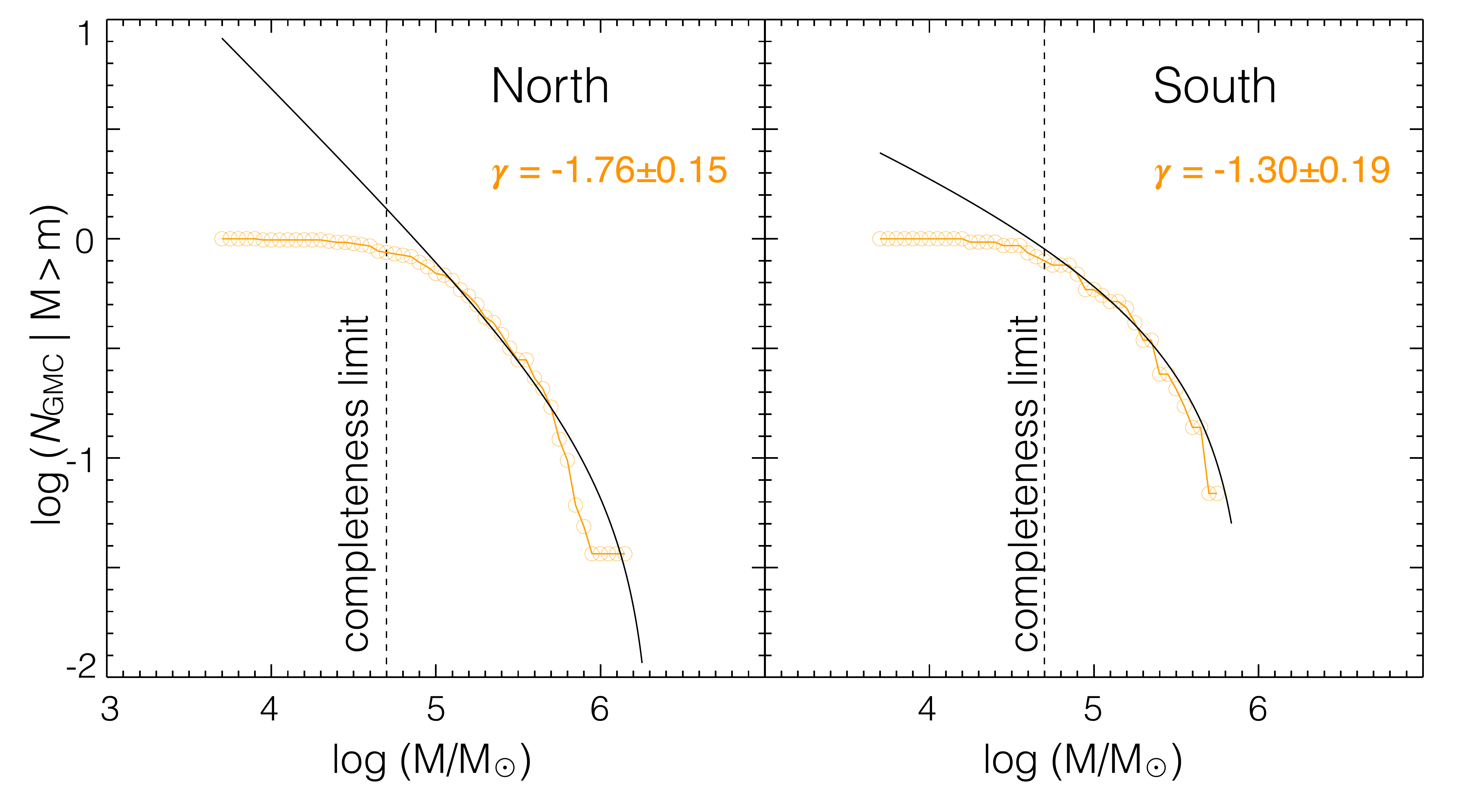}
\end{center}
\caption{Cumulative GMC mass distribution for the two environments in the TDG: north (left) and south (right). The vertical dashed lines indicate our completeness limit of $5 \times 10^4$\,M$_\odot$.
}
\label{fig:GMC_MS}
\end{figure}

The GMC mass spectrum is an indicator of GMC formation and dispersal timescales, modulated by mechanisms such as feedback and cloud-cloud collisions, and it is expected to depend on environment \citep{2015A&A...580A..49I,2015MNRAS.446.3608D,2015ApJ...806....7T,2017ApJ...836..175K}. Shallower GMC mass spectra imply a higher proportion of massive clouds within a given population; we denote the slope of the mass spectrum as $\gamma$.

\modif{Our} completeness limit of $5 \times 10^4$\,M$_\odot$
\modif{is} deeper than the PAWS survey in M51 \citep[$3.6 \times 10^5$\,M$_\odot$;][]{2014ApJ...784....3C}, but not as deep as the study of \citet{2018ApJ...857...19F} in NGC\,300 ($8 \times 10^3$\,M$_\odot$). \modif{This means that the smallest clouds may be unresolved and blending is likely for some.}

\modif{To compare with GMC mass spectra in other galaxies, here we focus on}
CPROPS, which is more widely used than SCIMES (shown in Appendix~\ref{sec:appendix}).
\modif{Unlike for GMC scaling relations, here we consider all the clouds identified by CPROPS (82 in the north and 29 in the south), as they all have extrapolated luminosities that permit to derive total masses ($M_{\rm GMC} = \alpha_\mathrm{CO}^{2-1} L_{\rm CO}$).}
Rather than a binned histogram (differential form), we fit a cumulative mass distribution, which has been argued to be more robust in the case of small samples \citep{2005PASP..117.1403R} and its use is more extended. When the cumulative GMC mass spectrum is well described by a power law, it can be expressed by the following formula:

\begin{equation}
N(M_{\rm GMC} > M) = \left(\frac{M}{M_0}\right)^{\gamma +1},
\end{equation}

\noindent
where $M_0$ is some reference mass (normalisation factor) and $\gamma$ is the slope of the power law.

Generally, the GMC mass spectrum can be better described as a truncated power law:

\begin{equation}
N(M_{\rm GMC} > M) = N_0 \left[ \left(\frac{M}{M_0}\right)^{\gamma +1} -1 \right],
\end{equation}

\noindent
where $N_0$ is the number of clouds more massive than $2^{1/(\gamma+1)} M_0$, the truncation mass where the distribution deviates from a power law.

We fitted both a truncated and non-truncated power law and found that the truncated version applies better to our data. Considering all GMCs in the TDG simultaneously \modif{(111 clouds)}, we find a slope of $\gamma = -1.76 \pm 0.13$. This value is compatible with the slope of the GMC mass spectrum in the inner Milky Way, where \citet{2016ApJ...822...52R} found $\gamma = -1.59 \pm 0.11$; for the first Galactic quadrant, \citet{2019MNRAS.483.4291C} also found a fully compatible value, $\gamma = -1.76 \pm 0.01$. Similar values have been found in M31 \citep[$\gamma = -1.63 \pm 0.2$;][]{2007ApJ...654..240R}; in M33 \citep[$\gamma \sim -1.6 \pm 0.2$;][]{2012A&A...542A.108G,2018A&A...612A..51B}; in the molecular ring and density-wave spiral arms in M51 \citep[$\gamma = -1.63 \pm 0.17$ to $\gamma = -1.79 \pm 0.09$;][]{2014ApJ...784....3C}; or in NGC\,300 \citep[$\gamma = -1.76 \pm 0.07$;][]{2018ApJ...857...19F}. Higher values have been reported in some cases: the outer Milky Way \citep[$\gamma = -2.1 \pm 0.2$;][]{2005PASP..117.1403R}; the LMC \citep[$\gamma = -2.33 \pm 0.16$;][]{2011ApJS..197...16W}; in the outer M33 \citep[$\gamma = -2.3 \pm 0.2$;][]{2012A&A...542A.108G}; in the material arm and the interarm region in M51 \citep[$\gamma = -2.44 \pm 0.40$ to $\gamma = -2.55 \pm 0.23$;][]{2014ApJ...784....3C}; in the lenticular galaxy NGC\,4526 \citep[$\gamma = -2.39 \pm 0.03$;][]{2015ApJ...803...16U}; and in the strongly barred spiral NGC\,1300 \citep[$\gamma = -2.20 \pm 0.04$;][]{2020MNRAS.493.5045M}. \citet{2017PASJ...69...18T} found a somewhat lower slope for GMCs in NGC\,1068, $\gamma = -1.25 \pm 0.07$, similar to the value derived by \citet{2014ApJ...784....3C} along the nuclear bar of M51 ($\gamma = -1.33 \pm 0.21$). All of these studies used the cumulative version of the mass spectrum, which should make the results comparable, but note that \citet{2012A&A...542A.108G} used a different algorithm to fit the cumulative mass distribution and, in general, differences in the detailed cloud segmentation strategies can affect the derived slopes.

We also checked the difference between the GMC mass spectrum in the quiescent north and in the star-forming south of the TDG. Interestingly, we found that the south has a shallower mass spectrum, $\gamma = -1.30 \pm 0.19$ instead of $\gamma = -1.76 \pm 0.15$. The difference is marginally significant given the large error bars (due to the relatively low number of clouds in each environment), but it points at an interesting idea: the higher proportion of more massive clouds in the south might be connected with the star formation activity in that region. This would be consistent with similar observations in star-forming rings or spiral arms in nearby galaxies, where shallower GMC mass spectra correlate with more intense star formation \citep{2014ApJ...784....3C,2017PASJ...69...18T}.

Another relevant quantity is the truncation mass of the GMC mass spectrum. Fitting all GMCs in the TDG,
we found a value of $M_0 = 1.9 \times 10^6$\,$M_\odot$. While M51 shows a significantly higher truncation mass (${\sim}2 \times 10^7$\,$M_\odot$; \citealt{2014ApJ...784....3C}), $M_0$ in the TDG is of the same order as the LMC, NGC\,4526, NGC\,300, or some Galactic results \citep[$M_0 \sim 1-4 \times 10^6$\,$M_\odot$;][]{2011ApJS..197...16W,2015ApJ...803...16U,2016ApJ...822...52R,2018ApJ...857...19F,2019MNRAS.483.4291C}.
These values agree with theoretical models for maximum cloud masses by \citet{2017MNRAS.469.1282R}.

\subsubsection{GMC scaling relations}
\label{Sec:Larson}

\begin{figure*}[t]
\begin{center}
\includegraphics[width=1.0\textwidth]{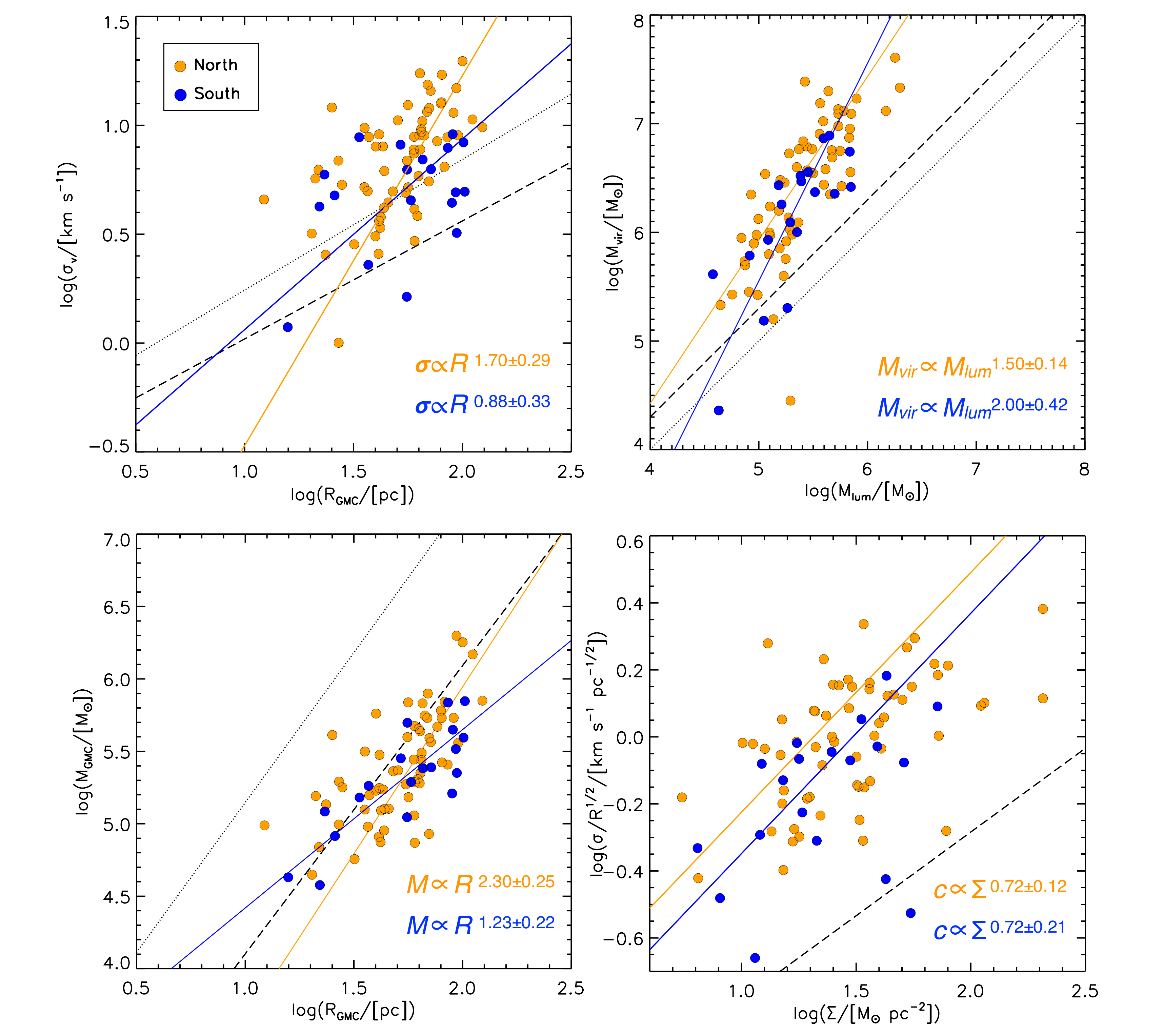}
\end{center}
\caption{Larson's relations for the ensemble of GMCs identified in the TDG using CPROPS. We differentiate between the quiescent environment in the north (orange points) and the star-forming region in the south (blue points). The solid lines represent the best ODR fits to the data (orange for north, blue for south). \textit{Top left panel:} size-linewidth relation. The dashed line is the relation for Galactic clouds from \citet{2001ApJ...551..852H} and the dotted line represents the best fit to extragalactic clouds from \citet{2008ApJ...686..948B}.  \textit{Top right panel:} virial mass-luminous mass relation. The dotted and dashed lines are the 1:1 and 2:1 relations, respectively. \textit{Bottom left panel:} size-mass relation. The dashed and dotted lines are the relations for Milky Way clouds from \citet{2010A&A...519L...7L} for an extinction threshold of $A_K=0.1$ and 1.5, respectively. \textit{Bottom right panel:}  Heyer plot showing $c=\sigma/\sqrt{R}$ as a function of the molecular gas mass surface density; the dashed line is the Galactic relation from \citet{2001ApJ...551..852H}.
}
\label{fig:Larson}
\end{figure*}

\begin{table*}[t!]
\begin{center}
\caption[h!]{Fits and rank coefficients for GMC scaling relations with CPROPS.}
\begin{tabular}{llccccccc}
\hline\hline 
\multirow{2}{*}{Relation}		& \multirow{2}{*}{Method}		&  \multicolumn{2}{c}{TDG} & \multicolumn{2}{c}{North}  & \multicolumn{2}{c}{South} &	 \multirow{2}{*}{MW  Slope}	\\
  \cmidrule(lr){3-4} \cmidrule(lr){5-6} \cmidrule(lr){7-8} 
& & Slope & $\rho$ & Slope & $\rho$ & Slope & $\rho$ &  \\
\hline 
\multirow{2}{*}{Size-linewidth}  & ODR & $2.00 \pm 0.33$ & \multirow{2}{*}{0.49 (0.00)} & $1.70 \pm 0.29$ & \multirow{2}{*}{0.62 (0.00)} & $0.88 \pm 0.33$ & \multirow{2}{*}{0.30 (0.21)} &	\multirow{2}{*}{0.5$^a$} \\
  & Bayesian & $1.69\substack{+0.57 \\ -0.39}$ &     & $1.50\substack{+0.55 \\ -0.37}$ &   & $0.86\substack{+2.59 \\ -0.80}$ &    &	 \\
 \hline
\multirow{2}{*}{Virial relation} & ODR & $1.66 \pm 0.13$ & \multirow{2}{*}{0.80 (0.00)} & $1.50 \pm 0.14$ & \multirow{2}{*}{0.81 (0.00)} & $2.00 \pm 0.42$ & \multirow{2}{*}{0.77 (0.00)} &	\multirow{2}{*}{0.8$^a$} \\
 & Bayesian & $1.59\substack{+0.22 \\  -0.18}$ &   & $1.43\substack{+0.26 \\ -0.20}$ &     & $1.70\substack{+3.47 \\ -1.19}$ &     &	 \\
 \hline
\multirow{2}{*}{Mass-size} & ODR & $2.08 \pm 0.19$ & \multirow{2}{*}{0.70 (0.00)} & $2.30 \pm 0.25$ & \multirow{2}{*}{0.68 (0.00)} & $1.23 \pm 0.22$ & \multirow{2}{*}{0.77 (0.00)} & \multirow{2}{*}{2.0$^a$} \\
 & Bayesian & $3.25\substack{+0.57 \\ -0.45}$ &     & $3.76\substack{+1.11 \\ -0.73}$ &  & $2.27\substack{+0.90 \\ -0.61}$ &     &  \\
 \hline
\multirow{2}{*}{Heyer plot} & ODR & $0.91 \pm 0.13$ & \multirow{2}{*}{0.46 (0.00)} & $ 0.72 \pm 0.12$ & \multirow{2}{*}{0.45 (0.12)} & $0.72 \pm 0.21$ & \multirow{2}{*}{0.44 (0.06)} &	\multirow{2}{*}{0.5$^b$} \\
 & Bayesian & $0.41\substack{+0.10 \\ -0.09}$ &       & $0.32\substack{+0.10 \\ -0.09}$ &  & $0.43\substack{+0.25 \\ -0.18}$ &    & \\
 \hline
\end{tabular}
\label{table:Larson}
\end{center}
\tablefoot{$\rho$ is the Spearman rank correlation coefficient, with the $p$-value indicated in parentheses. For each dataset we use two different fitting methods: orthogonal distance regression (ODR) and a Bayesian code, \texttt{BayesLineFit}. ($a$) \citet{1987ApJ...319..730S}. ($b$) \citet{2009ApJ...699.1092H}.}
\end{table*}

Here we focus on a widespread GMC diagnostics tool: the Larson scaling relations, shown in Fig.~\ref{fig:Larson}. \citet{1981MNRAS.194..809L} and \citet{1987ApJ...319..730S} found a tight correlation between the size and linewidth of GMCs, between their virial and luminous mass, and between their size and mass. Those seminal papers examined Galactic GMCs, but more recent work has expanded the analysis to extragalactic clouds as well. 
 We focus on CPROPS to minimise discrepancies with studies from the literature due to different segmentation codes. The best fits and Spearman rank correlation coefficients are listed in Table~\ref{table:Larson}. The linear regressions are obtained with two different methods. The first one is the popular orthogonal distance regression (ODR), where the data are fitted in the log-log plane using the Python implementation \texttt{scipy.odr}\footnote{https://docs.scipy.org/doc/scipy/reference/odr.html} that considers uncertainties in both variables simultaneously. As a complementary approach, we also fitted the observations using the publicly available Bayesian fitting code \texttt{BayesLineFit}\footnote{{\tt BayesLineFit} is available at http://astroweb.cwru.edu/SPARC/} \citep{2019MNRAS.484.3267L} \modif{that additionally considers the intrinsic scatter along the relation. \texttt{BayesLineFit} provides similar results as the ODR method but larger uncertainties (from exploring the full posterior distribution of the free parameters)}.
In both cases, these fits are based on the extrapolated and deconvolved GMC properties (see Sect.\,\ref{Sec:GMCproperties}), which result in substantially higher error bars; for instance, the median uncertainty on the cloud radius is 18\%, but as much as 55\% when considering extrapolation and deconvolution.

\paragraph{Size-linewidth relation}

A power-law relationship between size and velocity dispersion of GMCs  has been systematically found in the Milky Way, with a slope of $\sim$0.5, which has been interpreted as a manifestation of compressible turbulence within the molecular ISM 
\citep[e.g.][]{1987ApJ...319..730S,2001ApJ...551..852H,2016ApJ...822...52R,2017NJPh...19f5003K}. 
Observations of the Galactic central molecular zone imply a slightly steeper size-linewidth relation with a slope of $\sim$0.7 \citep{2012MNRAS.425..720S,2017A&A...603A..89K}.
The clouds in the TDG seem to have a moderate degree of correlation in this plane (left panel of Fig.\,\ref{fig:Larson}), but the slope $\gamma = 2.00 \pm 0.33$ is significantly steeper than the canonical Galactic value of $\sim$0.5. If we subdivide clouds into our two environments, we find a steeper slope for the north, but the difference is only marginally significant given the large error bars.

Within the extragalactic literature, there is some controversy as to whether a genuine size-linewidth relation holds. \citet{2008ApJ...686..948B} combined data from several nearby galaxies and found evidence for a size-linewidth relation ($\gamma = 0.60 \pm 0.10$).
On the other hand, a number of studies found that their data did not support the existence of a size-linewidth relation (e.g.\ \citealt{2012A&A...542A.108G} in M33; \citealt{2014ApJ...784....3C} in M51; \citealt{2013ApJ...779...46H} combining M33, the LMC, and M51)\modif{, or only weakly (e.g.\ \citealt{2018A&A...612A..51B} in M33)}. More recently, \citet{2018ApJ...857...19F} found a size-linewidth relation in NGC\,300 with a slope of $0.48 \pm 0.05$, with a larger dynamic range in GMC masses than the previous studies. 
Our data have a similar degree of correlation as found by \citet{2018ApJ...857...19F} in NGC\,300 (they quote a Pearson coefficient $r_P=0.55$). However, the substantially steeper slope of the size-linewidth relation in the TDG suggests that the organisation of the molecular medium into GMCs is different in this special environment.

\paragraph{Virial relation}

The second Larson relation suggests that GMCs are self-gravitating, since virial and luminous masses track each other. The top-right panel of Fig.\,\ref{fig:Larson} shows a strong correlation between virial and luminous masses. The global slope $1.66 \pm 0.13$ is considerably higher than the ${\sim}0.8{-}1$ value found in the Milky Way \citep{1987ApJ...319..730S} and in several nearby galaxies \citep[][]{2008ApJ...686..948B,2010MNRAS.406.2065H,2018ApJ...857...19F}. \citet{2014ApJ...784....3C} derived values for M51 that approach the one of the TDG ($\gamma \sim 1.3$ in the centre and spiral arms of M51, $\gamma \sim 1.5$ in the inter-arm region). Both the north and south of the TDG have similar slopes and a comparable degree of correlation.

Apart from the slope of the scaling relation, the virial masses in the TDG are on average ${\sim}1$\,dex higher than the luminous masses. Taken at face value, this result suggests that the most massive clouds are far from virial equilibrium and likely unbound, which is unusual for normal galaxies, but several factors could contribute to the discrepancy between $M_\mathrm{vir}$ and $M_\mathrm{lum}$. Firstly, the assumed distance of 14.5\,Mpc to Arp\,94 might be wrong, and this would affect the virial mass linearly but the luminous mass quadratically. Distance estimates to this system range within ${\sim}10{-}20$\,Mpc, with a few outliers at higher distances; even if the real distance to Arp\,94 is as much as 30\,Mpc, that could explain at most a factor of two offset in the $M_\mathrm{vir}-M_\mathrm{lum}$ relation, far from the factor of ten observed for the most massive clouds.
Another important source of uncertainty is the $\alpha_\mathrm{CO}$ conversion factor. A typical uncertainty of a factor ${\sim}2$ is expected \citep{2013ApJ...777....5S}; the near-solar metallicity of the TDG suggests that changes in $\alpha_\mathrm{CO}$ should not be much larger than that but, technically, we cannot rule out a variation of a factor ${\sim}10$ in $\alpha_\mathrm{CO}$ (as $\alpha_\mathrm{CO}$ has not been specifically characterised in TDGs); if this extremely high value of $\alpha_\mathrm{CO} \sim 40{-}50$\,M$_\odot$\,(K\,km\,s$^{-1}$\,pc$^2)^{-1}$ is true, that could fully account for the observed offset in $M_\mathrm{vir}-M_\mathrm{lum}$. What it would not be able to explain is the observed non-linear relation, implying that more massive clouds depart more severely from the virial estimate.
The cloud segmentation process could also introduce some systematic uncertainty, particularly for lower-mass clouds. However, the fact that the offset is especially pronounced for high-mass clouds makes segmentation issues an unlikely cause of the observed vertical offset. \modif{Finally, the overlap of GMCs along the line of sight could broaden linewidths, raising virial masses relative to the luminous masses.}

\begin{figure*}[t]
\begin{center}
\includegraphics[width=1.0\textwidth]{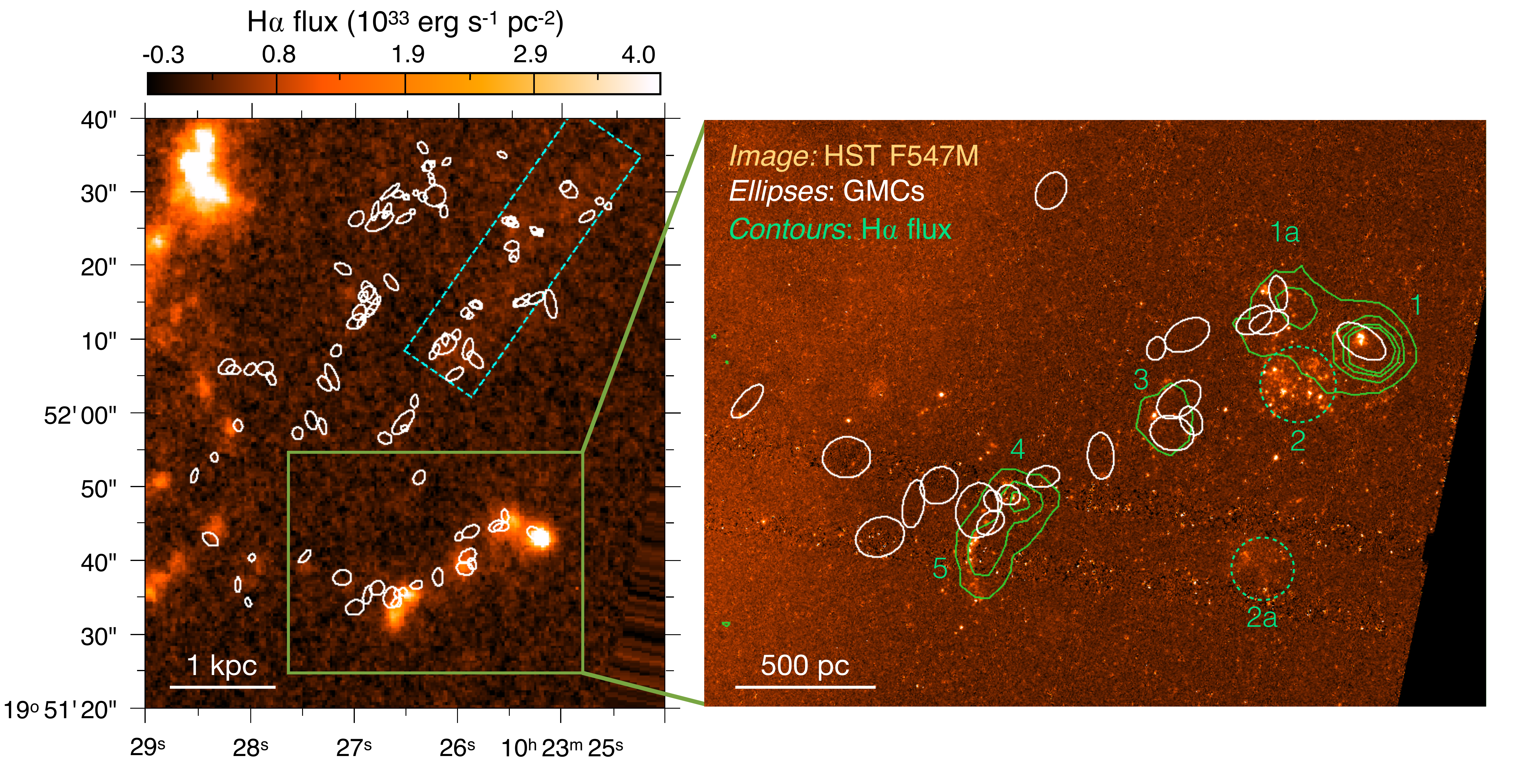}
\end{center}
\caption{\textit{Left panel:} H$\alpha$ image showing intense star formation in the south of the TDG, with white ellipses indicating the position, size, and orientation of GMCs identified with CPROPS; the cyan dashed rectangle marks an area of mildly enhanced H$\alpha$ emission as discussed in the text. \textit{Right panel:} HST WFC3/F547M image zoomed on the area marked in the left panel, revealing young stellar clusters in the south of the TDG. White ellipses indicate GMCs and green contours show constant H$\alpha$ flux (0.8, 2, 3, and $4 \times 10^{33}$~erg\,s$^{-1}$\,pc$^{-2}$). The numbering of regions follows \citet{2008ApJ...685..181L}; the dashed green regions are visible in $B$ band but lack detectable H$\alpha$ emission.
}
\label{fig:offsets}
\end{figure*}

In any case, it is physically plausible that clouds in the TDG are not (yet) in virial equilibirum between their kinetic energy and self-gravity, given that this is a newly formed galaxy that is in a dynamically complex context, where external pressure could play an important role \citep[see e.g.][]{2019ApJ...883....2S,2020ApJ...892..148S}. The most massive (larger) clouds tend to have higher velocity dispersions, making both the size-linewidth and virial relations super-linear.
\modif{We} also see some environmental variation in the virial relation within the TDG. The virial parameter (approximately equivalent to the ratio $M_\mathrm{vir}/M_\mathrm{lum}$) is higher for the clouds in the north (average ${\sim}13$) than for the clouds in the south (average ${\sim}8$), with significant overlap between both populations. 

\paragraph{Mass-size relation}

The third Larson relation describes an inverse correlation between size
and 
density, implying that all GMCs have approximately constant surface density.
The three Larson's relations are not independent from each other, and the applicability of two of Larson's relations immediately implies the third \citep[e.g.][]{2015ARA&A..53..583H}.
We find a statistically significant correlation between GMC mass and size in our data, with a global slope of $2.08 \pm 0.19$. A similar slope of ${\sim}2$ has been found in the Milky Way \citep{1987ApJ...319..730S}, in M51 \citep{2014ApJ...784....3C}, in the LMC \citep{2010MNRAS.406.2065H}, in combined data for nearby galaxies \citep{2008ApJ...686..948B}, or in NGC\,300 \citep{2018ApJ...857...19F}. Our observations are also in good agreement with the fit to Galactic data by \citet{2010A&A...519L...7L} for an extinction of $A_K = 0.1$, corresponding to a surface density of ${\sim}40$\,$M_\odot\,{\rm pc}^{-2}$.

\paragraph{Heyer plot}

Beyond Larson's relations, \citet{2009ApJ...699.1092H} introduced a plot condensing all three relations and highlighting systematic departures. If Larson's relations are strictly obeyed, data points should cluster around a point centred at $c=\sigma/\sqrt{R}=(\pi G \Sigma/5)^{1/2}$ for a constant value of $\Sigma$. However, similar to the findings from \citet{2009ApJ...699.1092H} in the Milky Way, the clouds in the TDG span an order of magnitude or more along each of the axes of the plot and show a positive correlation. For a given $\Sigma$, the points in the TDG are clearly displaced towards higher values of $c$, driven by the high velocity dispersions.

Given the variations in $\Sigma$ and $c$, and the moderate but significant positive correlation, our observations suggest that the velocity dispersion of a cloud depends both on the size of the GMC and its mass surface density, in agreement with 
\citet{2009ApJ...699.1092H}. Yet, at the same time, the vertical offset with respect to the Galactic observations suggests that GMCs in the TDG behave differently from those in the Milky Way, as also indicated by the departures from the first and second Larson's relations.

\subsection{Comparison between GMCs and star formation sites}
\label{Sec:offsets}

Star formation is predominantly happening in the south of the TDG, as shown by a variety of tracers in the optical and infrared, confirmed with spectroscopy \citep{2004ApJ...614..648M,2008ApJ...685..181L}. Fig.\,\ref{fig:offsets} shows archival H$\alpha$ and HST WFC3/F547M observations that clearly reveal emission from young stars formed in the south of the TDG. There are also some hints of diffuse but mildly enhanced H$\alpha$ emission towards the north-west of the TDG. At lower resolution, a similar flux enhancement is present in the \textit{Herschel} PACS 160\,$\mu$m map. When averaged inside a rectangle of $40'' \times 11''$ with PA$=55^\circ$ centred on RA=10:23:25.36 and Dec=+19:52:21.7 (cyan dashed line in Fig.\,\ref{fig:offsets}), the H$\alpha$ flux is ${\sim}1.5{-}2\times$ higher than in the immediate surroundings (mean H$\alpha$ flux of $2.2 \times 10^{32}$~erg\,s$^{-1}$\,pc$^{-2}$, as opposed to $1.4$ and $1.1 \times 10^{32}$~erg\,s$^{-1}$\,pc$^{-2}$ on adjacent rectangles towards the north or south, respectively). For reference, the H$\alpha$ contrast between the southern star-forming region and its surroundings (a factor of ${\sim}8$) is considerably higher than in the north-west. There are some scattered compact sources visible on the HST image towards the north of the TDG which might be young massive clusters \citep[see][for the efficient formation of clusters in TDGs]{2019A&A...628A..60F}. However, we also see similar knots towards the eastern side of NGC\,3227, far from the TDG, making it likely that these compact sources coincide in position with the TDG simply due to projection, something that we cannot test further without spectroscopy. 
In the rest of this section, we will focus on the predominant site of star formation in the south of the TDG.

Figure~\ref{fig:offsets} also identifies with green numbers the star-forming knots from \citet{2008ApJ...685..181L}. \modif{Most of these star-forming knots are overlapping with one or more GMCs, albeit there can be small projected offsets of about $0.8{-}1.6''=56{-}112$\,pc. Exceptions are represented by knots 2 and 5 that have no associated GMC (the nearest one is offset by $2{-}3'' = 140{-}210$\,pc).} Knot 2 was found to be the oldest in \citet{2008ApJ...685..181L} with ages above $\sim$100 Myr. This is consistent with the idea that young stars have had sufficient time to photodissociate or disperse the surrounding molecular gas; it would also explain why we no longer see H$\alpha$ emission. The clusters revealed by HST around knot 2 are very bright and cover a large area, reinforcing the idea that those young stars produce plenty of ionising radiation.

Finally, \citet{2008ApJ...685..181L} identified 2a as a region of stellar emission but without H$\alpha$. The HST imaging reveals weak emission throughout this area, but it is clearly less intense than the other knots. We did not detect any GMCs around this region (the closest GMC is at a projected distance of $7'' \sim 500$\,pc). It can well be that 2a is similar to knot 2 and the age is higher than the rest (${\sim} 100$\,Myr), such that there was enough time for any surrounding leftovers of molecular gas to get photodissociated or dispersed after the episode of star formation.

\section{Discussion}
\label{Sec:Discussion}

\subsection{A very large fraction of diffuse molecular gas}

\modif{In Sect.\,\ref{Sec:diffusefrac}, we found that as much as ${\sim}80$\% of the molecular emission in the TDG is extended, arising from scales of several kpc. This is likely tracing diffuse molecular gas, with a remarkably high contribution to the total CO budget. In fact,}
there is mounting evidence that nearby galaxies have a sizeable fraction of diffuse molecular emission, which is detected by single-dish telescopes but filtered out by interferometers. As opposed to compact molecular gas structures, this emission has been interpreted as evidence of thick molecular discs \citep[e.g.][]{2013ApJ...779...43P}. 
In principle, such extended emission could also be due to an ensemble of small clouds that are homogeneously distributed and separated by less than the synthesised beam of the interferometer \citep[e.g.][]{2013ApJ...779...43P,2015AJ....149...76C,2016AJ....151...34C}. \citet{1994ApJ...436L.173D} also found evidence for a thick molecular disc in the Milky Way (comparable in thickness to the \HI{} layer), and similar thick molecular discs have been reported for external galaxies, such as the edge-on spiral NGC\,891 \citep{1992A&A...266...21G} and M51 \citep[e.g.][]{2013ApJ...779...43P}. The fraction of diffuse gas found by these studies varies greatly from galaxy to galaxy, and is method-dependent, but typical values are $\sim$10-60\% \citep{2013ApJ...779...43P,2015AJ....149...76C,2020MNRAS.493.2872C,2020MNRAS.495.3840M}. It is noteworthy that \citet{2020MNRAS.495.3840M} found a particularly high fraction of diffuse molecular emission in the bar of NGC\,1300: 91\% in the region that they define as bar-A and 74\% in bar-B, as opposed to ${\sim}30{-}60$\% in the spiral arm and bar end.

In agreement with the findings in other galaxies, this diffuse CO component is associated with high velocity dispersions ($\sigma \sim 50{-}60$\,km\,s$^{-1}$ as opposed to $\sigma \lesssim 20{-}30$\,km\,s$^{-1}$ for the compact CO component at matched resolution; Fig.\,\ref{fig:diffuse_gas}). Interestingly, the position-velocity diagrams from Fig.\,\ref{fig:pvdiagram} generally show a very good agreement between this diffuse molecular component and atomic gas: even in regions of position-velocity space lacking compact molecular emission (GMCs), the diffuse molecular and atomic components seem to be well coupled. 

\modif{It is likely that} the peculiar and pristine nature of the TDG can explain this very high fraction of diffuse molecular emission. In 
the interacting system Arp\,94 where the TDG formed, the dynamical timescales are probably long, of the order of several 100\,Myr or even $\sim$Gyr (see Sect.\,\ref{Sec:dynamical} below). Therefore, 
atomic gas in 
the tidal tail \modif{may have} had enough time to form molecules on its own. This should be \modif{facilitated by}
the lack of strong UV emission from stars, unlike \modif{typical galaxy discs}
where radiation from young stars can easily photo-dissociate H$_2$.
In this framework, it seems plausible that molecular gas first appeared as a diffuse phase, and then
started to locally condense into more compact structures like the GMCs.

\subsection{Spatial variations in the \mbox{CO(2-1)/CO(1-0)} ratio}
\label{Sec:discussR21}

\modif{In Sect.\,\ref{Sec:R21}, we showed that the $R_{21}$=\mbox{CO(2-1)/CO(1-0)} ratio varies across the TDG with a mean value of $\sim$0.5.} Values of $0.5 {-} 1.0$ have typically been found in the Milky Way and at ${\sim}$kpc scales in the discs of nearby galaxies
\citep[e.g.][den Brok et al.\ submitted]{1994PASP..106.1112S,1996ApJ...460..334O,2010PASJ...62.1277Y,2009AJ....137.4670L}.
Therefore, the $R_{21}$ ratio found in 
this newborn TDG is not abnormal compared to nearby galaxies, even though the value of ${\sim} 0.35$ in the south lies at the lower end of what has been measured so far. 
\modif{The low $R_{21}$ measured might be connected to the nature of the interaction occurring in this system. }

Generally, low values of $R_{21}$ may indicate
that the gas is not in thermal equilibrium due to low volumetric densities, consistent with the high fraction of diffuse molecular gas that we inferred. This agrees with the recent study by \citet{2020MNRAS.495.3840M} in NGC\,1300, who found a very high fraction of diffuse molecular emission in its 
bar (74-91\%) mapping to a very low $R_{21}$ (0.17-0.34). We might be witnessing a similar effect in the TDG.

The observed gradient in $R_{21}$ in the TDG points to different physical conditions 
between the quiescent north and the star-forming south. Specifically, variations in the $R_{21}$ line ratio are expected to be driven by differences in the optical depth (closely related to density) \modif{and/or} the excitation temperature of the gas \citep{1994PASP..106.1112S, 2009AJ....137.4670L}.
\modif{We find} lower $R_{21}$ in the south, suggesting lower excitation temperature and/or lower density; these conditions are traditionally interpreted as associated with lower star formation efficiency. This might seem puzzling 
given that star formation is detected almost exclusively in the south
\modif{However,} the integrated $R_{21}$ value at $28''$ resolution is dominated by emission from diffuse molecular gas that is presumably not taking part in star formation. Based on synthetic observations of a simulated GMC, \citet{2017MNRAS.465.2277P} found that $R_{21}$ follows a bimodal distribution, with ${\sim}0.7$ for cold and dense parts and ${\sim}0.3$ in warmer and more diffuse regions. This is consistent with the fact that the CO emission 
in the TDG is dominated by diffuse molecular gas with a low $R_{21}$ value.

\subsection{Dynamical state and survival of the TDG}
\label{Sec:dynamical}

\modif{In Sect.\,\ref{Sec:kinematics}, we showed that} atomic and diffuse molecular gas \modif{as well as most GMCs} in the TDG display a north-south velocity gradient of ${\sim}100$\,km\,s$^{-1}$ (Fig.\,\ref{fig:pvdiagram}). To first order approximation, the diffuse CO component tracks atomic gas very well, suggesting that atomic and diffuse molecular gas are well mixed.
The \modif{north-south} gradient may be compatible with rotation, but given the uncertain geometry of the system it \modif{could} also \modif{reflect} stretching motions along the tidal tail. \modif{Moreover, there is some kinematic evidence for a gaseous ``bridge'' along the line of sight towards the North-East direction,} 
\modtwo{which is mostly made up of compact structures (GMCs) and connects the TDG with the background galaxy NGC\,3227.}

Overall, the velocity of the TDG along the line of sight is positive relative to the centre of NGC\,3227, \modif{similar} to the southern part of NGC\,3227. Since extinction arguments confirm that the TDG is currently in front of the spiral galaxy \citep{2004ApJ...614..648M}, \modif{the TDG must be} moving towards NGC\,3227, as long as the tangential velocities (which we cannot measure) are not much larger than the radial velocity. That ``backward'' motion towards NGC\,3227 would suggest that the TDG has been detached from its parent galaxy for a considerable amount of time; the timescales for such interactions are typically of the order of several hundred Myr or even ${\sim}1$\,Gyr.

\modif{It is unclear} whether the TDG will survive much \modif{longer.} 
Simulations show that condensations of matter at the tip of tidal tails have a larger probability of surviving long and eventually becoming detached from their parent system \modif{\citep[e.g.][]{2006A&A...456..481B}.} However, there is no way to robustly predict if a given matter condensation will survive, as this depends strongly on the dynamics of the system, which are not well known in detail for Arp\,94. 
In any case, 
\modif{we have used} J1023+1952 as a template laboratory to study what a young TDG can look like, independently from the future evolution of this specific system.

\subsection{GMC properties and Larson's relations}

\modif{In Sect.\,\ref{Sec:GMC}, we found that our TDG contain GMCs whose masses and sizes resemble those in spiral galaxies (Milky Way, M31, M33, M51, NGC\,300) and the LMC.}
This TDG, \modif{however, shows} distinct peculiarities compared to other galaxies. 
First of all, it has likely not settled into a disc and is probably not in dynamical equilibrium. Secondly, the fraction of diffuse molecular gas is extremely high. Thirdly, there are no stellar populations shining yet (neither young nor old) across most of the area covered by the TDG. 
\modif{This} implies that the processes determining \modif{the masses and sizes of GMCs}
are not necessarily associated with disc dynamics or stellar feedback, since those do not play an important role in our TDG. In other words, 
regular disc rotation or stellar feedback do not seem to be essential for GMCs to have certain sizes, masses, and to follow a mass distribution which approaches a truncated power law with a slope of $\gamma \sim -1.8$.

We have examined Larson's relations and found that these clouds 
show important departures from the scaling relations observed in the Milky Way and other galaxies. While clouds seem to obey the same size-mass relation observed in other galaxies, virial masses almost always exceed luminous masses, often by as much as an order of magnitude. 
This might be an observational artifact, given the large uncertainty in $\alpha_\mathrm{CO}$ \modif{in TDGs}. If real, the offset could indicate that clouds are super-virial and have not settled into dynamical equilibrium yet. \modif{The higher velocity dispersion of the most massive clouds} explains the super-linear trend in the size-linewidth and virial relations, steeper than observed in the Milky Way and other galaxies. \modif{This could be driven by the blending of clouds with slightly different velocities inside a beam.}

\subsection{Offsets between GMCs and young stellar clusters}
\label{Sec:discussoffsets}

\modif{In Sect.\,\ref{Sec:offsets}}, we studied the \modif{spatial} relation between young stellar clusters and GMCs in the star-forming part of the TDG. Broadly speaking, we found three different cases: 1) good spatial agreement between the position of GMCs and young clusters; 2) small but noticeable offsets between clusters and GMCs, typically of 50-100 pc; and 3) clusters not associated with any GMCs, the nearest GMC \modif{being} at a projected distance of $\gtrsim$200\,pc \modif{(keeping in mind that the true physical separation might be larger)}. Interestingly, the smallest offsets coincide with the youngest star-forming knots, while the largest offsets are associated with the oldest knots, consistent with an age trend. 

Our results agree with the different situations identified by \citet{2009ApJS..184....1K} in the LMC (see also \citealt{1999PASJ...51..745F}) or the findings by \citet{2012ApJ...761...37M} and \citet{2012A&A...542A.108G} in M33, \citet{2018ApJ...863L..21K} in NGC\,628,  \citet{2018MNRAS.481.1016G} in NGC\,7793, and \citet{2019MNRAS.483.4707G} in M51, \modif{with typical offsets ${\sim}40{-}100$\,pc} if present.
In particular, \citet{2018MNRAS.481.1016G, 2019MNRAS.483.4707G} found a clear trend between cluster age and the GMC-cluster separation, emphasising the idea that the offsets result from the time evolution of individual star-forming regions. 
In the framework of the uncertainty principle for star formation \citep{2014MNRAS.439.3239K,2018MNRAS.479.1866K}, \citet{2020MNRAS.493.2872C} derived region separation lengths between molecular gas and star formation peaks ($\lambda$) in the range $107{-}267$\,pc for nine nearby star-forming galaxies. Under the assumption of complete spatial randomness, this separation length translates into an expectation value of ${\approx}0.44 \lambda$ for the nearest neighbour distance \citep[see eq.\ 9 in][]{2019Natur.569..519K}, which results in an expected separation of ${\sim}50{-}120$\,pc between emission peaks, in good agreement with the range of observed offsets.
Comparable offsets between star-forming sites and GMCs have been found using alternative tracers of star formation \citep[e.g.\ infrared emission or radio continuum;][]{2010ApJ...721.1206C, 2014ApJ...795..156W, 2017A&A...601A.146C, 2019A&A...625A..19Q}.

The observed offsets between GMCs and young clusters suggest that, once it is triggered, star formation in the TDG proceeds in a very similar fashion as in standard dwarf and spiral galaxies. Moreover, the fact that the offsets are also quantitatively similar to other galaxies points to the idea that the dispersal timescales of molecular gas around a young star-forming region is fairly universal, \modif{independent} from the peculiar nature of the TDG.

\subsection{What drives star formation in TDGs?}
\label{Sec:SFtriggering}

\begin{figure}[t]
\begin{center}
\includegraphics[trim=0 0 0 0, clip,width=0.48\textwidth]{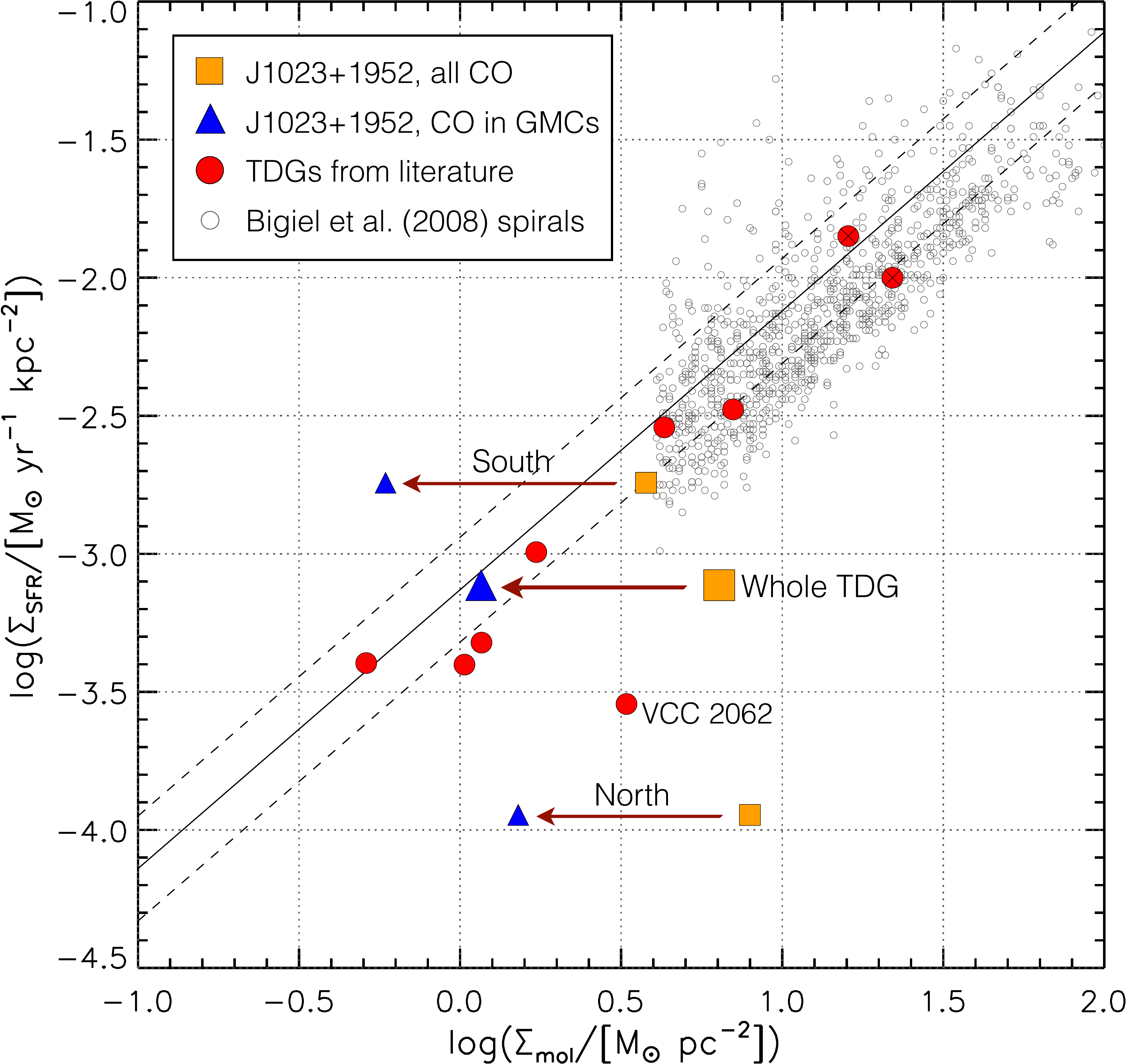}
\end{center}
\caption{Molecular Kennicutt-Schmidt relation for 
J1023+1952 compared to
TDGs
from the literature (compiled by \citealt{2016A&A...590A..92L}). The orange squares represent our 
measurements of the total molecular gas, whereas the blue triangles correspond to the flux in GMCs identified with CPROPS, associated with compact molecular emission. The small squares and triangles differentiate the north from the south of the TDG (as defined in Fig.\,\ref{fig:mosaic}). For reference, the open gray circles in the background show the kpc-scale measurements in spiral galaxies from \citet{2008AJ....136.2846B}; the solid line represents their best fit to that sample and the dashed line its standard deviation. The molecular gas literature data for TDGs comes from single-dish telescopes except for the circles marked with an additional cross.
}
\label{fig:KSplot}
\end{figure}

\modif{To place our work within a broader context, Fig.\,\ref{fig:KSplot} shows} the molecular Kennicutt-Schmidt relation \citep{1959ApJ...129..243S,1998ApJ...498..541K} \modif{for spiral galaxies and other TDGs}.
When considering all the molecular gas, our TDG seems globally inefficient at forming stars, similarly to \modif{the nearby TDG} VCC\,2062 \citep{2016A&A...590A..92L}. The inefficiency is particularly acute in the north, \modif{while}
the south 
is similar to other TDGs \modif{and the extrapolation of spiral galaxies to lower surface densities}.

Figure~\ref{fig:KSplot} also highlights how much the presence of extended CO emission can affect the apparent efficiency of star formation on the Kennicutt-Schmidt plot. For J1023+1952, there is a considerable difference depending on whether we focus on compact structures revealed 
identified as GMCs 
or if we account for all the CO flux.
For the entire system, the star formation activity seems normal when considering the compact molecular gas in GMCs, but it appears as highly inefficient at forming stars if we also include extended emission.
When inspecting the north and the south separately, we find that the star formation surface density in the north is below the value expected from the Kennicutt-Schmidt relation both for the total and compact molecular gas surface density. Conversely, the star formation activity in the south follows the Kennicutt-Schmidt relation and is even above the relation when examining the compact molecular emission.

\modif{As suggested by \citet{2008ApJ...685..181L},}
increased large-scale turbulence might be preventing gas from forming stars in the north of the TDG. Our ALMA observations 
\modif{confirm that} the increased velocity dispersion in the north arises primarily from a higher velocity dispersion in the diffuse gas and a larger relative velocity among \modif{GMCs}. Cloud-cloud collisions have \modif{also} been invoked to explain the triggering of star formation \citep[e.g.][]{2000ApJ...536..173T,2014MNRAS.445L..65F,2015MNRAS.446.3608D,2015ApJ...806....7T}. However, if clouds are similarly extended along the line of sight across the TDG, the lower velocity dispersion among clouds in the south would suggest a lower cloud-cloud collision rate, making this factor less likely as an explanation for the enhanced star formation.

The distribution of star forming sites along a ridge \modif{in the south}, with an age gradient in the stellar clusters, could be suggestive of some propagation mechanism, such as stochastic self-propagating star formation \citep{1976ApJ...210..670M,1978ApJ...223..129G,1995MNRAS.275..507S}. If a given mechanism initiates star formation in a certain cloud complex, feedback from star formation could subsequently trigger the collapse of other clouds in the neighbourhood, with a certain time delay. 
\modif{Several studies} have claimed to detect the imprint of stochastic self-propagating star formation in observations of 
dwarf and irregular galaxies \citep{1981A&A....98..371F,1985ApJ...297..599D,1987ApJ...317..190M,2001AJ....121.1024N}.

\modif{We found that the} global gas surface density (atomic+molecular) is higher in the north, with a higher molecular-to-atomic ratio, and a larger cloud-to-cloud velocity dispersion. 
\modif{In this context, the lack of strong star formation in the north could be explained as an increased star formation threshold due to the higher gas density and turbulence.}
This would qualitatively agree with the predictions from \citet{2018ApJ...854...16E}, and with observations of a lower star formation efficiency of the (dense) molecular gas towards the centre of the Milky Way (\citealt{2013MNRAS.429..987L,2014MNRAS.440.3370K,2020arXiv200705023B}) and the centres of other nearby galaxies \citep{2015AJ....150..115U,2016ApJ...822L..26B,2018ApJ...858...90G,2019A&A...625A..19Q}. In general, wherever molecular gas surface densities are higher, it is harder for the gas to turn into stars because the ``barrier'' for star formation increases. This interpretation also agrees with the observation of a higher molecular-to-atomic gas ratio in the north, which would be expected if local gas density and pressure are higher.
\modif{In any case,} the largest limitation is that we are dealing with projected surface densities, and 
\modif{the observed} surface densities might not necessarily track the volumetric gas densities that presumably modulate the threshold of star formation.

Alternatively (or in addition) to differences in star formation thresholds, 
\modtwo{the triggering of star formation in the south and the inhibition of star formation in the north may be the result of complex dynamical mechanisms, such as tidal torques, stretching motions, converging flows, ram pressure, and shear \citep[e.g.][]{2009ApJ...706...67R,2013MNRAS.436..839S,2015MNRAS.447.2512P,2018MNRAS.474..580P}.}
\modtwo{At any rate,}
\modif{our results support a scenario where dynamical effects associated with the interaction, and not a lack of GMCs in the north, explain why star formation is limited to the south of the TDG.}

\section{Summary and conclusions}
\label{Sec:conclusions}

We have presented new ALMA observations of \mbox{CO(2-1)} emission in J1023+1952, a TDG in the system Arp 94, with $0.64'' \sim 45$\,pc resolution. To our knowledge, these are the highest resolution observations of molecular gas in a TDG so far, and the first time that individual GMCs are resolved in a TDG. Our main findings can be summarised as follows:

\begin{enumerate} 

  \item We find an extremely high fraction of extended molecular emission (${\sim}80{-}90$\%), which is filtered out by the ALMA interferometer but recovered by the total power antennas. This emission is arising from scales larger than ${\sim}2$\,kpc and is likely tracing diffuse molecular gas. High fractions of extended molecular emission have been found in nearby galaxies, but this is clearly at the upper end of what has been observed so far.

  \item For our assumed distance (14.5\,Mpc) and $\alpha_\mathrm{CO}$ conversion factor ($\alpha_\mathrm{CO}^{2-1} = 4.4\,M_\odot$\,(K\,km\,s$^{-1}$\,pc$^2)^{-1}$), when measured at matched $6.3''$ resolution, the total molecular gas mass in the TDG is $8.6 \times 10^7$\,M$_\odot$, very similar to the total amount of atomic gas ($8.4 \times 10^7$\,M$_\odot$). This results in a very high molecular-to-atomic gas ratio ($\sim 1$), with significant variations between north ($\sim 1.5$) and south ($\sim 0.4$).

  \item The global \mbox{CO(2-1)/CO(1-0)} ratio is quite low ($R_{21}=0.52$), at the lower end of measurements in nearby galaxies, and this becomes particularly extreme towards the south of the TDG ($R_{21} = 0.38$). For the western part of NGC\,3227 covered by our ALMA setup, the average $R_{21}=0.53$ is similar to the TDG, suggesting that the low $R_{21}$ value could be \modtwo{related to the interaction}.
  
  \item We identified 111 GMCs in the TDG with CPROPS. These GMCs are spread throughout the entire TDG, and not preferentially concentrated in the star-forming part. They show similar sizes, masses, and linewidths as GMCs in the Milky Way and in other nearby galaxies ($R \sim 10{-}100$\,pc, $M_\mathrm{lum} \sim 10^4-10^6$\,M$_\odot$, and $\sigma_v$ of a few km\,s$^{-1}$). The total molecular gas in GMCs adds up to 3$\times 10^7$\,$M_\odot$ (extrapolated to perfect sensitivity), which is 18\% of the total molecular mass (1.63$\times 10^8$\,$M_\odot$ in the ALMA cube at native resolution, including total power, and without any masks).

  \item GMCs in the TDG follow a mass spectrum that is well described by a truncated power law with slope $\gamma = -1.76 \pm 0.13$ and truncation mass of $M_0 = 1.9 \times 10^6$\,$M_\odot$. These values are comparable to the findings in the Milky Way and other nearby galaxies, in spite of the very different conditions of the TDG.
  
  \item GMCs in the TDG follow scaling relations with similarly strong degrees of correlation as in the Milky Way and other galaxies. The slope of the size-mass relation is in agreement with previous observations. However, the size-linewidth and virial relation in the TDG are super-linear and significantly steeper than observed elsewhere. The high velocity dispersion of clouds in the TDG, particularly extreme for the most massive clouds, can explain the observed super-linear relations. Virial masses almost always exceed luminous masses, often by an order of magnitude (especially in the north). If this is not an observational artifact (e.g.\ due to changes in $\alpha_{\rm CO}$), it could indicate that clouds are super-virial.
  
  \item We find varying spatial offsets between young stellar clusters and GMCs in the star-forming part of the TDG. The smallest offsets ($\lesssim$50\,pc) correspond to the youngest star-forming knots, while the largest offsets ($\gtrsim$200\,pc) are associated with the oldest knots, consistent with an age trend.
  
\end{enumerate}

In conclusion, our observations highlight the complex organisation of molecular gas in a TDG, with a large reservoir of diffuse molecular gas and many clouds that are not forming stars. Since the main difference between north and south is kinematic, we suggest that clouds in the north are stabilised against collapse by dynamical effects. GMCs in this system have a similar mass distribution and sizes as observed in other galaxies, but they show departures from the scaling relations involving velocity dispersion, and this is particularly acute in the quiescent north. On the other hand, the south of the TDG suggests that, once the clouds begin to form stars, the process of star formation and subsequent feedback proceeds in a very similar fashion as in other galaxies. \modif{This process might be assisted by a mechanism such as stochastic self-propagating star formation.}

\small  
\begin{acknowledgements}   
We appreciate helpful comments from Bruce Elmegreen and Andreas Schruba.
MQ and SGB acknowledge support from the research project  PID2019-106027GA-C44 from the Spanish Ministerio de Ciencia e Innovaci\'on.
UL acknowledges support by the research
project  AYA2017-84897-P from the Spanish Ministerio
de Econom\'ia y Competitividad, from the European Regional Development Funds (FEDER), and the Junta de Andaluc\'ia (Spain) grants FQM108.
FB acknowledges funding from the European Research Council (ERC) under the European Union’s Horizon 2020 research and innovation programme (grant agreement No. 726384/EMPIRE).
SGB acknowledges support through grants PGC2018-094671-B-I00 and AYA2016-76682-C3-2-P (MCIU/AEI/FEDER,UE).
JMDK gratefully acknowledges funding from the Deutsche Forschungsgemeinschaft (DFG, German Research Foundation) through an Emmy Noether Research Group (grant number KR4801/1-1) and the DFG Sachbeihilfe (grant number KR4801/2-1), as well as from the European Research Council (ERC) under the European Union's Horizon 2020 research and innovation programme via the ERC Starting Grant MUSTANG (grant agreement number 714907).
\end{acknowledgements}

\bibliography{mq.bib}{}
\bibliographystyle{aa}{}

\appendix
\section{Alternative identification of GMCs with SCIMES}
\label{sec:appendix}

\begin{figure*}[t]
\begin{center}
\includegraphics[width=0.68\textwidth]{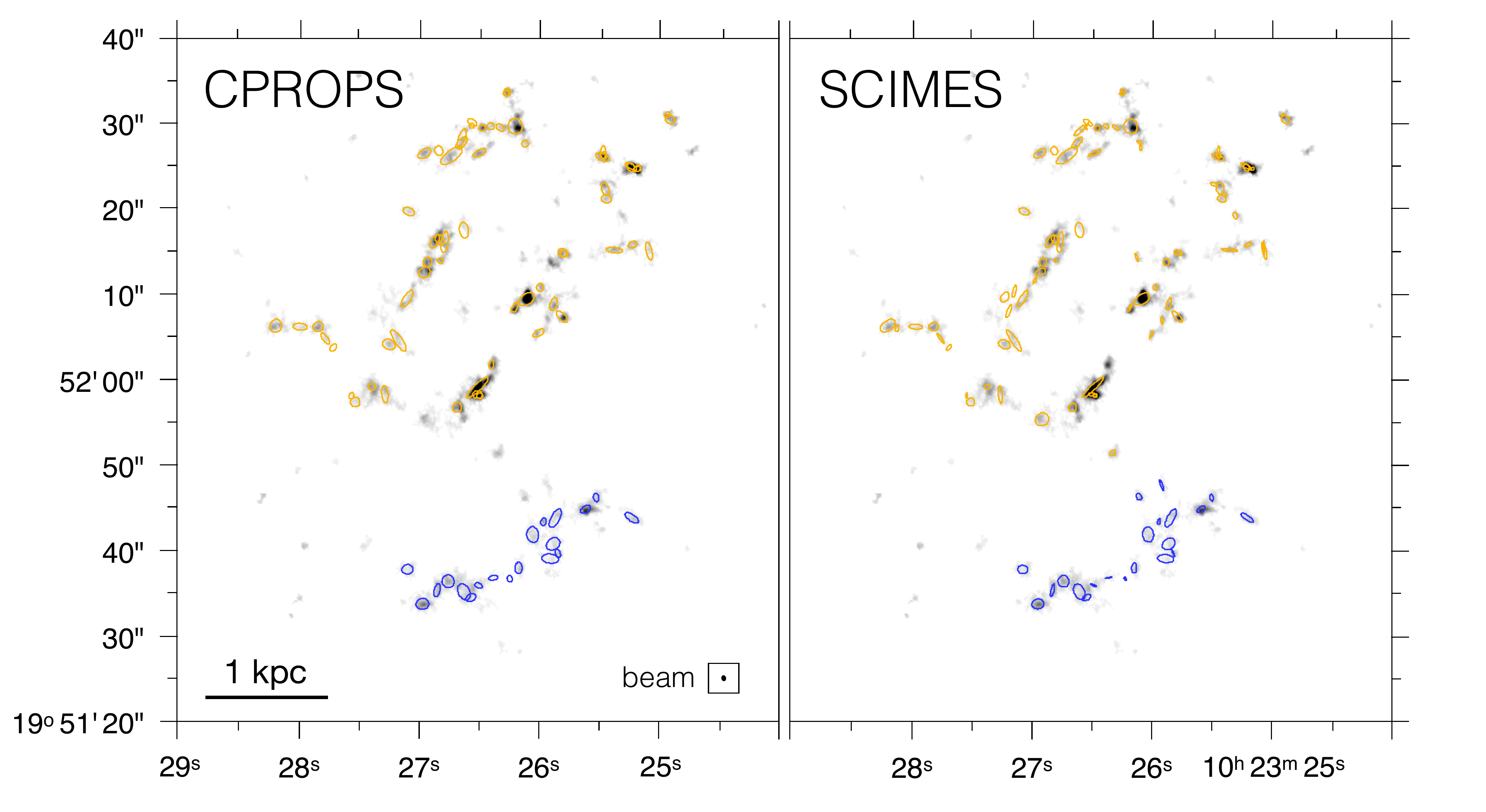}
\end{center}
\caption{GMC positions and orientations for CPROPS (\textit{left panel}) compared to SCIMES (\textit{right panel}) indicated on the ALMA \mbox{CO(2-1)} integrated intensity map as in Fig.\,\ref{fig:GMC_map}.
GMCs are shown as ellipses with the extrapolated and deconvolved major and minor axes (2nd moment of emission). Orange and blue ellipses represent clouds in the north and south of the TDG, respectively. The bottom-right corner of the left panel shows the ALMA synthesised beam ($0.69'' \times 0.60''$ with PA=11$^\circ$).
}
\label{fig:CPROPS-SCIMES}
\end{figure*}

We applied SCIMES \citep{2015MNRAS.454.2067C} to our data as an alternative approach to identify GMCs. SCIMES is based on dendrograms, tree diagrams that reflect the hierarchical structure of a dataset \citet{2008ApJ...679.1338R}. A dendrogram tree is made up of trunks (top hierarchical level), and substructures such as branches (intermediate level) and leaves (lowest level): a branch can split into multiple leaves. \modif{The goal of applying SCIMES to our data is two-fold: we want to check if our results are sensitive to the specific choice of code and examine the degree of hierarchy within the molecular medium. }

\subsection{GMC segmentation with SCIMES}
\label{Sec:SCIMES}

SCIMES relies on a graph-based clustering technique (spectral clustering) to optimally segment the structure tree based exclusively on the input data.
The method is tuned to the segmentation of the molecular ISM by employing the luminosity and volume of the isosurfaces to establish the similarity matrices that are key to the spectral clustering.

SCIMES \modif{has been} used to isolate molecular clouds and cloud substructures in Galactic data (e.g.\ filaments or cores). For extragalactic data, however, given the limited spatial resolution, leaves typically map to entire GMCs while branches can correspond to larger structures, such as groups of GMCs, that are physically connected.

\begin{table*}[t!]
\begin{center}
\caption[h!]{Properties of structures derived by SCIMES.}
\setlength{\tabcolsep}{5pt} 
\begin{tabular}{lcccccccccc}
\hline\hline
\multirow{2}{*}{Property} & \multirow{2}{*}{Unit}  & \multicolumn{3}{c}{TDG}  & \multicolumn{3}{c}{north} & \multicolumn{3}{c}{south} \\
\cmidrule(lr){3-5} \cmidrule(lr){6-8} \cmidrule(lr){9-11} & & min  & median & max  & min  & median & max & min  & median & max \\
\hline
{\bf Leaves} &  &   &   &   &   &   &   &   &   &   \\
$T_{\rm max}$ & K &  0.3 &  0.4 &  1.1 &  0.3 &  0.4 &  0.9 &  0.3 &  0.5 &  1.1 \\
$L_{\rm CO}$ & K\,km\,s$^{-1}$\,pc$^{2}$ &   5.2$\times 10^3$ &   4.1$\times 10^4$ &   5.0$\times 10^5$ &   5.2$\times 10^3$ &   4.2$\times 10^4$ &   5.0$\times 10^5$ &   6.5$\times 10^3$ &   4.1$\times 10^4$ &   1.8$\times 10^5$ \\
$R$ & pc &   14.6 &   51.4 &  122.8 &   14.6 &   51.3 &  122.8 &   18.0 &   55.8 &  102.5 \\
$\sigma_v$ & km\,s$^{-1}$ &    1.0 &    6.1 &   19.7 &    1.0 &    6.5 &   19.7 &    1.2 &    4.7 &    9.8 \\
$M_{\rm lum}$ & $\mathrm{M_\odot}$ &   2.3$\times 10^4$ &   1.8$\times 10^5$ &   2.2$\times 10^6$ &   2.3$\times 10^4$ &   1.9$\times 10^5$ &   2.2$\times 10^6$ &   2.9$\times 10^4$ &   1.8$\times 10^5$ &   7.8$\times 10^5$ \\
$M_{\rm vir}$ & $\mathrm{M_\odot}$ &   2.8$\times 10^4$ &   2.4$\times 10^6$ &   4.1$\times 10^7$ &   5.9$\times 10^4$ &   2.7$\times 10^6$ &   4.1$\times 10^7$ &   2.8$\times 10^4$ &   2.2$\times 10^6$ &   7.6$\times 10^6$ \\
\hline
{\bf Branches} &  &   &   &   &   &   &   &   &   &   \\
$T_{\rm max}$ & K &  0.4 &  0.6 &  1.1 &  0.4 &  0.5 &  0.8 &  0.5 &  0.6 &  1.1 \\
$L_{\rm CO}$ & K\,km\,s$^{-1}$\,pc$^{2}$ &   2.5$\times 10^4$ &   2.2$\times 10^5$ &   7.7$\times 10^5$ &   2.5$\times 10^4$ &   1.9$\times 10^5$ &   7.7$\times 10^5$ &   1.1$\times 10^5$ &   2.9$\times 10^5$ &   6.1$\times 10^5$ \\
$R$ & pc &   14.7 &   97.9 &  310.2 &   14.7 &   95.7 &  181.4 &   81.7 &  153.2 &  310.2 \\
$\sigma_v$ & km\,s$^{-1}$ &    4.3 &   12.6 &   26.2 &    4.3 &   12.6 &   26.2 &    4.3 &   10.6 &   13.5 \\
$M_{\rm lum}$ & $\mathrm{M_\odot}$ &   1.1$\times 10^5$ &   9.7$\times 10^5$ &   3.4$\times 10^6$ &   1.1$\times 10^5$ &   8.5$\times 10^5$ &   3.4$\times 10^6$ &   4.9$\times 10^5$ &   1.3$\times 10^6$ &   2.7$\times 10^6$ \\
$M_{\rm vir}$ & $\mathrm{M_\odot}$ &   3.2$\times 10^5$ &   1.6$\times 10^7$ &   1.3$\times 10^8$ &   3.2$\times 10^5$ &   1.6$\times 10^7$ &   1.3$\times 10^8$ &   2.5$\times 10^6$ &   1.3$\times 10^7$ &   5.3$\times 10^7$ \\
\hline
{\bf Trunks} &  &   &   &   &   &   &   &   &   &   \\
$T_{\rm max}$ & K &  0.3 &  0.6 &  1.1 &  0.3 &  0.5 &  0.9 &  0.5 &  0.7 &  1.1 \\
$L_{\rm CO}$ & K\,km\,s$^{-1}$\,pc$^{2}$ &   4.3$\times 10^4$ &   2.0$\times 10^5$ &   1.1$\times 10^6$ &   4.3$\times 10^4$ &   2.0$\times 10^5$ &   1.1$\times 10^6$ &   5.7$\times 10^4$ &   5.9$\times 10^5$ &   6.2$\times 10^5$ \\
$R$ & pc &   36.3 &  108.1 &  307.9 &   36.3 &  108.0 &  208.1 &   76.6 &  211.6 &  307.9 \\
$\sigma_v$ & km\,s$^{-1}$ &    4.3 &   12.8 &   29.4 &    7.7 &   13.4 &   29.4 &    4.3 &   12.8 &   14.3 \\
$M_{\rm lum}$ & $\mathrm{M_\odot}$ &   1.9$\times 10^5$ &   8.8$\times 10^5$ &   4.9$\times 10^6$ &   1.9$\times 10^5$ &   8.8$\times 10^5$ &   4.9$\times 10^6$ &   2.5$\times 10^5$ &   2.6$\times 10^6$ &   2.7$\times 10^6$ \\
$M_{\rm vir}$ & $\mathrm{M_\odot}$ &   1.5$\times 10^6$ &   2.1$\times 10^7$ &   1.9$\times 10^8$ &   3.8$\times 10^6$ &   2.1$\times 10^7$ &   1.9$\times 10^8$ &   1.5$\times 10^6$ &   4.5$\times 10^7$ &   5.2$\times 10^7$ \\
\hline
\end{tabular}
\label{table:GMCpropsSCIMES}
\end{center}
\end{table*}

\subsection{Results with SCIMES}

The results from this section confirm that GMC properties do not look very different when identified with CPROPS or SCIMES. Fig.\,\ref{fig:CPROPS-SCIMES} \modif{shows that,} even though the segmentation is not identical on a cloud-by-cloud basis, the overall distribution of GMCs is similar, and clouds also show similar sizes and axis ratios. This is further emphasised by Table\,\ref{table:GMCpropsSCIMES}, where the properties of the `leaves' can be contrasted against the typical properties of GMCs from CPROPS as listed in Table\,\ref{table:GMCproperties}.

\modif{As} an ensemble CPROPS and SCIMES yield cloud populations with similar properties, with a small trend for SCIMES to recover slightly more luminous and massive clouds. There are also small differences in the GMC mass spectrum: the slope of the global truncated power-law fit is slightly higher with SCIMES, $\gamma = -1.84 \pm 0.15$ (instead of $\gamma = -1.76 \pm 0.13$), but they are well compatible within the uncertainties. The mass spectrum in the north also remains steeper with SCIMES ($\gamma = -1.74 \pm 0.16$ vs $\gamma = -1.57 \pm 0.37$), but the difference between north and south becomes even less significant than with CPROPS.

The distribution of molecular mass in GMCs between north and south is also comparable to what we found with CPROPS: 2.9$\times 10^7\,M_\odot$ in the north and 6.8$\times 10^6\,M_\odot$ in the south (as opposed to 2.4$\times 10^7\,M_\odot$ and 5.8$\times 10^6\,M_\odot$ with CPROPS). The GMCs identified with SCIMES also behave very similarly in terms of the scaling relations that we examined in Sect.\,\ref{Sec:Larson}. Table\,\ref{table:LarsonSCIMES} shows the Spearman rank correlation coefficients and the best power-law fits for the GMC scaling relations with SCIMES. The correlation coefficients are generally similar to the ones implied by CPROPS, and the fitted relations also tend to agree with the results from CPROPS within the error bars; the strongest departures take place in cases where the uncertainty of the fits was already large (e.g.\ size-linewidth relation for the south). The super-linear nature of the virial relation persists with SCIMES.

\modif{The} application of SCIMES to our data has revealed some hierarchical clustering in the molecular gas of the TDG. SCIMES identified a total of 127 leaves, 49 branches, and 26 trunks in the TDG. 
On average, each trunk has 3.9 leaves pending directly from \modif{it} and 1.9 branches (but some of these branches are indirect, in the sense that a branch can split into other branches). Each branch, in turn, hosts 3 leaves on average. There is a total of 26 leaves that do not have a direct parent.

In the Milky Way, where much higher spatial resolutions can be achieved, GMCs have been proven to consist of hierarchical substructures (filaments, clumps, etc.), but we cannot resolve such intra-GMC substructures here. It can happen, conversely, that the GMCs are not fully isolated from each other and are actually embedded in larger macrostructures. \modif{SCIMES} suggests that this is the case to some extent, but \modif{not to the extreme of} all GMCs in the TDG \modif{being} connected \modif{up to a single} `tree'.
The application of SCIMES to data in external galaxies is so far limited and the interpretation of these hierarchical macrostructures is not straight-forward when the synthesized beam is comparable to an entire GMCs.

\clearpage
\onecolumn
 \begin{landscape}
 
\normalsize

\begin{table*}[t!]
\begin{center}
\caption[h!]{Fits and rank coefficients for GMC scaling relations with SCIMES.}
\begin{tabular}{llccccccc}
\hline\hline 
\multirow{2}{*}{Relation}		& \multirow{2}{*}{Method}		&  \multicolumn{2}{c}{TDG} & \multicolumn{2}{c}{North}  & \multicolumn{2}{c}{South} &	 \multirow{2}{*}{MW  Slope}	\\
  \cmidrule(lr){3-4} \cmidrule(lr){5-6} \cmidrule(lr){7-8} 
& & Slope & $\rho$ & Slope & $\rho$ & Slope & $\rho$ &  \\
\hline 
\multirow{2}{*}{Size-linewidth}  & ODR & $1.40 \pm 0.20$ & \multirow{2}{*}{0.46 (0.00)} & $1.41 \pm 0.20$ & \multirow{2}{*}{0.57 (0.00)} & $0.03 \pm 0.14$ & \multirow{2}{*}{0.07 (0.77)} &	\multirow{2}{*}{0.5$^a$} \\
& Bayesian & $1.27\substack{+0.53 \\ -0.36}$ &     & $1.33\substack{+0.45 \\ -0.33}$ &   & $0.01\substack{+0.27 \\ -0.32}$ &    &	 \\
 \hline
\multirow{2}{*}{Virial relation} & ODR & $1.51 \pm 0.11$ & \multirow{2}{*}{0.80 (0.00)} & $1.42 \pm 0.11$ & \multirow{2}{*}{0.83 (0.00)} & $1.41 \pm 0.35$ & \multirow{2}{*}{0.66 (0.00)} &	\multirow{2}{*}{0.8$^a$} \\
 & Bayesian & $1.42\substack{+0.19 \\  -0.16}$ &   & $1.35\substack{+0.21 \\ -0.17}$ &     & $1.20\substack{+3.39 \\ -1.49}$ &     &	 \\
 \hline
\multirow{2}{*}{Mass-size} & ODR & $2.02 \pm 0.16$ & \multirow{2}{*}{0.71 (0.00)} & $2.22 \pm 0.21$ & \multirow{2}{*}{0.70 (0.00)} & $1.21 \pm 0.20$ & \multirow{2}{*}{0.76 (0.00)} & \multirow{2}{*}{2.0$^a$} \\
 & Bayesian & $3.08\substack{+0.48 \\ -0.37}$ &     & $3.54\substack{+0.81 \\ -0.57}$ &  & $2.22\substack{+0.80 \\ -0.58}$ &     &  \\
\hline
\multirow{2}{*}{Heyer plot} & ODR & $0.72 \pm 0.08$ & \multirow{2}{*}{0.54 (0.00)} & $ 0.59 \pm 0.08$ & \multirow{2}{*}{0.52 (0.00)} & $0.72 \pm 0.16$ & \multirow{2}{*}{0.58 (0.01)} &	\multirow{2}{*}{0.5$^b$} \\
 & Bayesian & $0.37\substack{+0.08 \\ -0.08}$ &       & $0.30\substack{+0.08 \\ -0.08}$ &  & $0.45\substack{+0.25 \\ -0.18}$ &    & \\
 \hline
\end{tabular}
\label{table:LarsonSCIMES}
\end{center}
\tablefoot{$\rho$ is the Spearman rank correlation coefficient, with the $p$-value indicated in parentheses. For each dataset we use two different fitting methods: orthogonal distance regression (ODR) and a Bayesian code, \texttt{BayesLineFit}. ($a$) \citet{1987ApJ...319..730S}. ($b$) \citet{2009ApJ...699.1092H}.}
\end{table*}

\begin{table*}[t!]
\begin{center}
\caption[h!]{GMC catalogue from CPROPS. The first ten rows are shown for guidance on the format. The full table is available in electronic format.}
\setlength{\tabcolsep}{5pt} 
\begin{tabular}{cccccc
r@{\hspace{0.5\tabcolsep}}C{0.2cm}@{\hspace{0.5\tabcolsep}}l
r@{\hspace{0.5\tabcolsep}}C{0.2cm}@{\hspace{0.5\tabcolsep}}l
c
r@{\hspace{0.5\tabcolsep}}C{0.2cm}@{\hspace{0.5\tabcolsep}}l
r@{\hspace{0.5\tabcolsep}}C{0.2cm}@{\hspace{0.5\tabcolsep}}l
r@{\hspace{0.5\tabcolsep}}C{0.2cm}@{\hspace{0.5\tabcolsep}}l
c}
\hline\hline 
 Cloud ID & RA$\rm_{J2000}$ & DEC$\rm_{J2000}$ & $(a/b)$ & PA & $v_{\rm cen}$ & \multicolumn{3}{c}{$R$}  & \multicolumn{3}{c}{$\sigma_{\rm v}$} & $T_{\rm peak}$ & \multicolumn{3}{c}{$L_{\rm CO}$} & \multicolumn{3}{c}{$M_{\rm lum}$} & \multicolumn{3}{c}{$M_{\rm vir}$} & Environment \\
  & (deg) & (deg)  &  & (deg) & (km\,s$^{-1}$) & \multicolumn{3}{c}{(pc)} & \multicolumn{3}{c}{(km\,s$^{-1}$)} & {(K)}  & \multicolumn{3}{c}{(K\,km\,s$^{-1}$\,pc$^2$)} & \multicolumn{3}{c}{($10^3\,\mathrm{M_\odot}$)} & \multicolumn{3}{c}{($10^3\,\mathrm{M_\odot}$)} &  \\
   \hline
1	&	155.8609	&	19.8751	&	3.58	&	33.0	&	1039.9	&		& &		&	5.6	& $\pm$ & 	3.8	&	0.394	&	12710	& $\pm$ & 	9534	&	56	& $\pm$ & 	42	&		& &		&	N	\\
2	&	155.8606	&	19.8750	&	1.77	&	146.5	&	1044.6	&	45.7	& $\pm$ & 	23.3	&	4.4	& $\pm$ & 	2.1	&	0.431	&	28853	& $\pm$ & 	10756	&	127	& $\pm$ & 	47	&	931	& $\pm$ & 	987	&	N	\\
3	&	155.8610	&	19.8751	&	3.68	&	87.8	&	1048.8	&		& &		&	1.7	& $\pm$ & 	2.5	&	0.394	&	9822	& $\pm$ & 	8829	&	43	& $\pm$ & 	39	&		& &		&	N	\\
4	&	155.8606	&	19.8749	&	1.09	&	69.3	&	1063.0	&	12.3	& $\pm$ & 	20.3	&	4.6	& $\pm$ & 	1.3	&	0.396	&	22143	& $\pm$ & 	6205	&	97	& $\pm$ & 	27	&	266	& $\pm$ & 	478	&	N	\\
5	&	155.8625	&	19.8702	&	1.90	&	52.0	&	1066.6	&		& &		&	9.1	& $\pm$ & 	3.6	&	0.394	&	31495	& $\pm$ & 	11951	&	139	& $\pm$ & 	53	&		& &		&	N	\\
6	&	155.8619	&	19.8712	&	2.05	&	58.0	&	1077.5	&	69.2	& $\pm$ & 	15.2	&	15.4	& $\pm$ & 	2.2	&	0.547	&	180161	& $\pm$ & 	25269	&	793	& $\pm$ & 	111	&	17012	& $\pm$ & 	6101	&	N	\\
7	&	155.8622	&	19.8705	&	1.13	&	60.9	&	1089.5	&	60.0	& $\pm$ & 	16.4	&	8.9	& $\pm$ & 	1.9	&	0.582	&	107533	& $\pm$ & 	15746	&	473	& $\pm$ & 	69	&	4892	& $\pm$ & 	2607	&	N	\\
8	&	155.8623	&	19.8702	&	1.22	&	39.6	&	1135.3	&	79.6	& $\pm$ & 	13.5	&	12.6	& $\pm$ & 	1.5	&	0.455	&	137245	& $\pm$ & 	12763	&	604	& $\pm$ & 	56	&	13120	& $\pm$ & 	4050	&	N	\\
9	&	155.8612	&	19.8749	&	1.73	&	29.6	&	1067.2	&		& &		&	1.6	& $\pm$ & 	2.4	&	0.375	&	18177	& $\pm$ & 	9285	&	80	& $\pm$ & 	41	&		& &		&	N	\\
... &  ... &   ... &  ... &  ... &  ... &   & ... &   &    & ... &    &   ... &   & ... &  &    & ... &   &   & ... &   &   ... \\
  \hline
\end{tabular}
\label{table:results}
\end{center}
\end{table*}

\end{landscape}

\end{document}